\documentclass[aps,prl,groupedaddress,twocolumn,showkeys]{revtex4-1}
\usepackage{graphicx} 
\usepackage{color} 
\usepackage{hyperref}
\usepackage{amssymb}
\usepackage{cases} 
\usepackage{varioref}
\usepackage{ulem}

\begin{document}


\title{Self-consistent quantum field theory for the
  characterization of complex random media by short laser pulses}


\author{Andreas Lubatsch}
\affiliation{Physikalisches Institut, Rheinische Friedrich-Wilhelms
  Universit\"at Bonn, Wegelerstr. 8, 53115 Bonn,  Germany\\ Georg-Simon-Ohm University of Applied Sciences, Ke{\ss}lerplatz
  12, 90489 N\"urnberg, Germany}
\author{Regine Frank}
\email[]{regine.frank@googlemail.com, regine.frank@rutgers.edu}
\affiliation{Department of Physics and Astronomy, Serin Physics Laboratory, Rutgers University, 136 Frelinghuysen Road, Piscataway, NJ
  08854-8019, USA\\ Physikalisches Institut, Rheinische Friedrich-Wilhelms
  Universit\"at Bonn, Wegelerstr. 8, 53115 Bonn, Germany}


\date{\today}

\begin{abstract}
We present a quantum field theoretical method for the
  characterization of disordered complex media with short laser pulses in an optical coherence tomography setup (OCT). We
  solve this scheme of coherent transport in space and time with weighted
  essentially non-oscillatory methods (WENO). WENO is preferentially used for the
  determination of highly non-linear and discontinuous processes including
  interference effects and phase transitions like Anderson localization of light. The theory
  determines spatio-temporal characteristics of the scattering mean free path
  and the transmission cross section that are directly measurable in
  time-of-flight (ToF) and pump-probe 
  experiments. The results are a measure of the coherence of multiple
  scattering photons in passive as well as in optically soft random
  media. Our theoretical results of ToF are instructive in spectral regions where material characteristics such as the scattering mean
  free path and the diffusion coefficient are methodologically almost
  insensitive to gain or absorption and to higher-order nonlinear
  effects. Our method is applicable to OCT and other
  advanced spectroscopy setups including samples of strongly scattering  mono- and polydisperse complex nano-
  and microresonators.
\end{abstract}

\keywords{wave propagation in random media, coherence tomography, Anderson
  localization, transmission and
  absorption, time resolved light scattering spectroscopy, glasses, porous materials; granular materials in materials
  science, complex systems}


\maketitle


\section{\label{sec:intro}Introduction}

\begin{figure} 
\scalebox{0.65}{\includegraphics{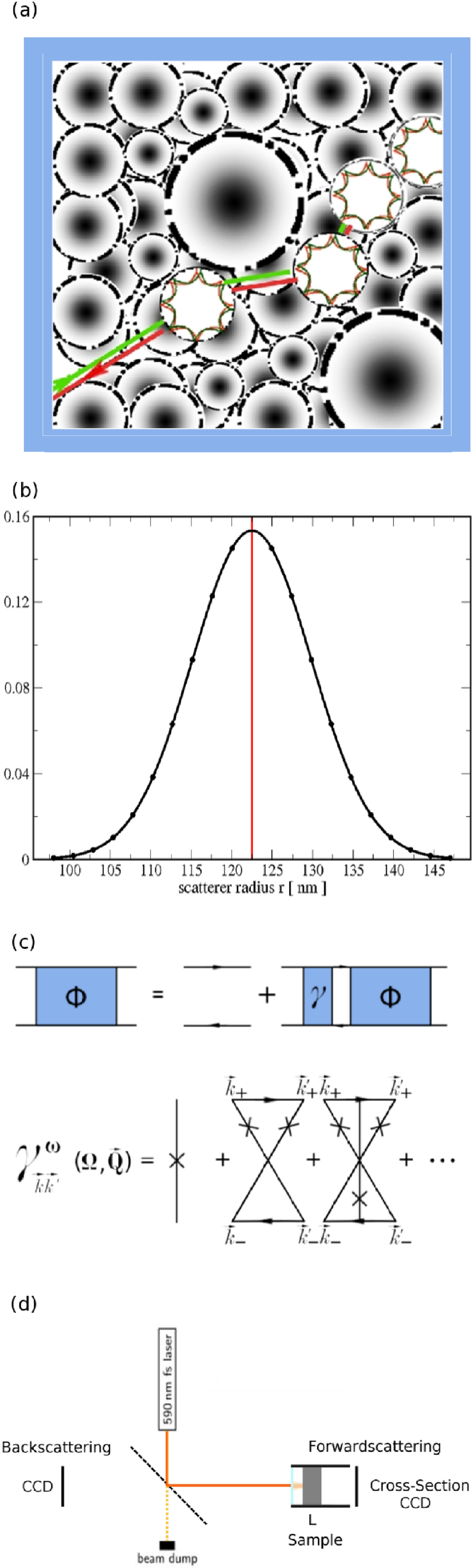}}
\caption{(a) Characterization of disordered random media, mono- and
  polydisperse Mie scatterers. (b) Gaussian distribution of
  scatterers radii. (c) Quantum field theoretical approach, self-consistent
  Bethe-Salpeter equation including the Cooperon. (d) Experimental
setup, forward-  and backscattering. The discussion is found in the {\it introduction} \ref{sec:intro}.}
\label{Setup}
\end{figure}

The characterization of disordered media has been fascinating the community
ever since, and groundbreaking analysis methods like coherent backscattering (CBS)
\cite{Akkermans}, dynamic light scattering (DLS)
\cite{Pecora,Goodman,Provencher,Schurtenberger,Scheffold,Baravian}, diffusing wave
spectroscopy (DWS) \cite{Pine} and optical
coherence tomography (OCT) \cite{OCT,Drex,OCTIEEE,Zhang,Zhou,Dogariu2} have been developed on the basis of
transport of classical electromagnetic waves in random media and photonic crystals
\cite{Akkermans,John1,John2}. Classical methods in the time (TD-OCT)  and in
the frequency domain (SD-OCT) have
been generalized with great success for polarization-sensitive optical
coherence tomography \cite{deBoer, Yamanari, Gosh,Gompf1}, for applications in metrology
\cite{Serpo1,Serpo2,Dogariu,Gompf} as well as for opto-medical imaging
\cite{Vellekop_Aegerterimaging,Mosk,Lippok}. Sub- and hyperdiffusive random media \cite{Isichenko} have attracted great
interest. Quantum-optical coherence tomography (QOCT) using
entangled-photon-sources in a Hong-Ou-Mandel interferometer \cite{Mandel,Ou} has been
demonstrated \cite{Teich1,Teich2,Teich3}. It yielded an improvement of the
resolution of a factor of two compared to OCT. Probing the submicron scale characteristics
of transport of light is a crucial aspect in the understanding of dynamic properties of disordered
random media \cite{Amon,Wuensche,Torres}.  Many of these approaches however do not
account for multiple scattering at all, thus self-interference effects which
can yield up to a factor of two enhancement of coherently scattered light with
respect to the incoming intensity in
passive scatterers ensembles \cite{Akkermans}, the CBS peak, are neglected. 
All these methods have in common that a
systematic incorporation of multiple scattering processes of light by
optically soft scatterers, so the decoherence of light due to light-matter
interaction, absorption, all orders of non-linear processes \cite{Novitsky} as
well as scattering losses, are not incorporated in a systematic way
\cite{Baravian,Tearney,Knuttel,Zvyagin}. The range from weakly to strongly
scattering non-conserving media so far is
not covered by a systematic methodology
\cite{Alfano,Nieuwenhuizen,Rotter2,Rotter} while quantitative fluorescence
spectroscopy (QFS) in the time and the frequency regime is broadly investigated in turbid media and soft matter \cite{Mycek}. Technological applications in solid
state physics and soft matter such as novel light sources, random lasers and
solar cells based on multiple scattering in disordered
arrangements of active resonators may profit from such novel techniques \cite{CaoPNAS,Chu,Sun,Erden}. Non-invasive and
non-destructive methods of optical analysis and medical imaging that can
detect reflected signals as small as $10^{-10}$ of the incident intensity and beyond
\cite{OCT,OCTIEEE,Zhang,Zhou,Dogariu2} might be improved again by orders of magnitude. Their
application range could be systematically improved in the fields of dynamic and
non-conserving media, nematic liquid crystals, semiconductors for telecom
applications, glasses and tissue \cite{Amon,Boccara,Vignolini,Garcia,Scheffold2}.\\
In this article we develop a quantum field theory for photonic transport in
dense multiple scattering complex random media, see Fig. \ref{Setup}(a). The
Bethe-Salpeter equation, which governs the propagation of the intensity \cite{FrankPRB2006,Frank2011}, is solved for a propagating short laser
pulse in random media with the help of a weighted essentially
non-oscillatory solver (WENO) \cite{Shu,Harten,Osher,Liu,Shu2020,Lax} in the space and time dependent
framework \cite{Warburton}, Fig. \ref{Setup}(c). We
are  going beyond the diffusion approximation by including interferences
and repeated self-interferences in
the sense of all orders of maximally crossed diagrams, the Cooperon \cite{VW1,VW2}. This is
generally associated with the Anderson transition of electrons as well as of
light, sound and
matter waves in
multiple scattering random media
\cite{Anderson,Chabanov,PhysicsToday,Fishman,SegevChristodoulides,Genack2,Shapiro1,Shapiro2,SAspect,Sanchez2019,HuS,Maret2012,WS,NPHOT2013},
and thus with quantum effects \cite{VW1,VW2}. The random medium is assumed to be
optically complex and we are including non-linearities, absorption and gain, which in consequence request the
implementation of suitable conservation laws by means of the Ward-Takahashi
identity for non-conserving media and resonators. The random medium can be comprised of ensembles of
arbitrarily shaped particles as well as correlated disorder and glassy systems in principle \cite{Vardeny}. We focus in this
work on independent  non-conserving Mie scatterers \cite{Genack,DUALSYM} in
strongly scattering ensembles of a high filling fraction, for instance mono- and polydisperse
complex TiO$_2$ powders \cite{Evans,Chakravarthy}. Such ensembles are well
known for showing a pronounced Mie signature in their transport characteristics such as
the scattering mean free path as it has been determined experimentally also by
coherent backscattering for optically passive systems \cite{Alfano,Akkermans,John2}, Fig. \ref{Setup}(d). \\
We derive in what follows self-consistent results for
{\it conserving} and for {\it non-conserving} random media and  we show that
absorption and gain or non-linearities can be characterized in a
time-of-flight experiment (ToF) by the fraction of transmitted photons which experience a
delay due to coherent multiple scattering. It will be shown that the deviation in the long time limit from the pure diffusive
case provides fundamental knowledge about the subtle nature of the scattering
ensemble and it's complexity in the sense of the resonator properties of the
single scatterer. 

Fig1SETUP5.pdf

\section{Quantum field theory for multiple scattering of photons}
\label{QFT}

\subsection{Nonlinear response}

The electrodynamics for transport of light in random media is described basically  by the wave equation
\begin{eqnarray}
\Delta \vec E - \frac{1}{c^2}\frac{\partial^2 \vec E}{\partial t^2} =
\frac{1}{c^2\epsilon_0}\frac{\partial^2\vec P}{\partial t^2}.
\label{NL}
\end{eqnarray}
where the polarizability in bulk matter $\vec P(\vec E)$ can be decomposed in linear
and non-linear part 
\begin{eqnarray}
\vec P &=& \epsilon_0 [\chi^{(1)} \vec E + (\chi^{(2)} \vec E)\vec
E + ((\chi^{(3)} \vec E)\vec E)\vec E +\,...]\\
&=& \vec P_{L}\, +\, \vec P_{NL}.\nonumber
\end{eqnarray}
As opposed to the non-linear Schr\"odinger equation for matter waves
\cite{Sanchez-PLewenstein}, in the presented quantum field theoretical formalism  the wave
equation allows for the
straightforward incorporation of the Mie resonance \cite{Mie} as a  classical
geometrical effect in the sense of a whispering gallery resonance of the light wave
at the inner surface of the complex scatterer \cite{Lubatsch05}. In general the polarizability is defined as $\vec P(\vec E) =
\epsilon_0(\epsilon_1 -1)\vec E + \vec P_{NL}$. $\epsilon_0$ is the dielectric
constant in free space, $\epsilon$ is
defined in the literature as the material specific dielectric coefficient $\epsilon\,=\,1\,+\,\chi$, $c$ is the speed of light. Higher-order processes, for
instance Kerr
media \cite{Novitsky} with the dielectric susceptibility $\chi^{(2)}$ \cite{Evans,Chakravarthy}, are in electrodynamics
classified by the dependency of $\vec P$ to the electrical field $\vec E$
\cite{Wegener} without loss of generality. It is well
known that both the conductivity as well as the susceptibility contribute to
the permittivity, so in general ${\rm Im} \epsilon\,=\,{\rm Im}(\chi)\,+\,{\rm
  Re}(\sigma/\omega)$ is given. Absorption and optical gain are represented by a finite positive or negative
imaginary part of the dielectric
function, so in general ${\rm Im} \epsilon \ne 0 $ is assumed. We take into
account the Mie scatterer Fig.(\ref{Setup}) for the determination of the single particle self-energy
contribution $\Sigma^\omega_{\vec
  k}$ of the quantum field-theoretical approach in what follows. The Mie scattering
coefficients of n-th order are written as \cite{Mie,Bohren}
\begin{eqnarray} 
a_n &=&
\frac{
    m\Psi_n(my)\Psi_n^{\prime}(y)
     - \Psi_n(y)\Psi_n^{\prime}(my)
}
{
      m\Psi_n(my)\xi_n^{\prime}(y)
     -   \xi_n(y)\Psi_n^{\prime}(my)}\\\nonumber
\\\nonumber	
b_n &=&
\frac{
    \Psi_n(my)\Psi_n^{\prime}(y)
     -m\Psi_n(y)\Psi_n^{\prime}(my)
}
{
    \Psi_n(my)\xi_n^{\prime}(y)
     -m\xi_n(y)\Psi_n^{\prime}(my)
}.
\end{eqnarray} 

where $m$ here denotes the complex refractive index, $y\,=\,2\pi\,r_{scat}/\lambda$ is
the size parameter, $\lambda$ is the wavelength of light and $r_{scat}$ is the
spheres radius. Prime denotes the derivative with respect to the argument of
the function. $\Psi_n$ and $\xi_n$ are Riccati-Bessel functions defined in
terms of the spherical Bessel function  and in terms of the Hankel function
\cite{Mie,Bohren}. The characteristics of the active scatterer and the active
embedding matrix are
described by ${\rm Im} \epsilon_{scat}$ and   ${\rm Im} \epsilon_b$ in the
following.

\subsection{Self-consistent Bethe-Salpeter equation for inelastic multiple scattering of photons}

For the theoretical description we may use any distribution and shape of particles which
may be described in the form of a scattering matrix. Here we consider in our results monodisperse spheres as well as a Gaussian distribution of
spherical scatterers located at random positions
\cite{Stoerzer1,Stoerzer2,Buehrer,Maret2012,WS,NPHOT2013,WiersmaPRL2007,WiersmaPRA2008,WiersmaLevi2012,WiersmaFibonacci2005,WiersmaLAGPRE1996},
see Fig. (\ref{Setup}). The scatterers and the background medium are described by the dielectric
constants $\epsilon_{scat}$ and $\epsilon_b$, respectively. In this work we use
unpolarized light and therefore we consider the {\it scalar} wave equation which has been Fourier transformed from time $t$ to frequency $\omega$ and reads
\begin{equation} 
\label{eq:field} 
\frac{\omega^2}{c^2} \, \epsilon (\vec{r}\,) \Psi_\omega(\vec{r}\,) 
+ \nabla ^2 \Psi_\omega( \vec{r}\,) 
= -i \omega \frac{4\pi}{c^2}  j_\omega(\vec{r}\,)\ , 
\end{equation} 
where $c$ denotes the vacuum speed of light  
and  $j_\omega (\vec{r}\,)$ the current. The current $j_\omega (\vec{r}\,)$
may be expanded in orders of $\omega$ \cite{Kubo1,Peterson}. We do not take
into account here a coupling to a microscopic model for
dynamical feedback of the optically driven crystal in the non-equilibrium
\cite{LubatschEPJB,LubatschSymmetry,LubatschAPPLSCI2020}, or to chaos modulations and chaotic systems \cite{Reichl}. The dielectric constant is spatially dependent,
$ \epsilon(\vec{r}\,) = \epsilon_b + \Delta\epsilon\, V(\vec{r}\,)$, and
the dielectric contrast is defined as
$\Delta\epsilon = \epsilon_{scat} - \epsilon_b$. The dielectric contrast
describes in principal the arrangement of scatterers through the function 
$V(\vec{r}\,) = \sum_{\vec{R}} S_{\vec{R}}\,(\vec{r}-\vec{R}\,)$, with 
$S_{\vec{R}}\,(\vec{r}\,)$ a localized shape function 
at random locations $\vec{R}$.

The intensity is related to the field-field correlation function
$\langle \Psi(\vec{r},\,t\,) \Psi^*(\vec{r}\,',t\,'\,)\rangle$, where  angular
brackets $\langle\ldots\rangle$ denote the
ensemble or disorder  average. To calculate the  field-field correlation the Green's
function formalism is used, the  (single-particle) Green's function is 
related to the (scalar) electrical field by
\begin{equation} 
\label{SP_Green_function_field} 
 \Psi(\vec{r},\,t\,) = 
\left\lmoustache  \!  {\rm d}^3r\,'    \right.\!\!
\left\lmoustache  \!  {\rm d}t\,'    \right.
G( \vec{r} \, ,\vec{r} \,'\, ;\,t\,,t'\,)      j(\vec{r}\,'\,,t'\,)\,.
\end{equation}
The Fourier transform  of the retarded, disorder averaged single-particle 
Green's function of Eq. (\ref{eq:field}) reads,
\begin{equation}
\label{SP_Green_function}
G_{\vec{q}}^{\omega} = \frac{1}
{\epsilon_b (\omega /c)^2 - \vert \vec{q} \vert^2 - \Sigma^{\omega}_{\vec{q}}} \ ,
\end{equation} 
where the retarded self-energy $\Sigma _{\vec{q}} ^{\omega}$  
arises from scattering of the random potential 
$-(\omega /c)^2(\epsilon_{scat} - \epsilon_b )V(\vec{r}\,)$. Using Green's
functions the mode density, the local density of photonic states (LDOS),
$N(\omega)$ may be expressed as  $N(\omega)=-(\omega/\pi) {\rm Im} G_0^{\omega}$,
where we use the abbreviation $G_0^{\omega}\equiv \int d^3q/(2\pi)^3\,
G_{\vec{q}}^{\omega}$. 
We study the transport of the already introduced field-field correlation by
considering the four-point correlator,  defined in terms of the non-averaged
Green's functions  $\hat G$, $\hat G^*$ in momentum and frequency space  as 
$\Phi^{\omega}_{\vec{q}\vec{q}^{\prime}\,\,}(\vec{Q},\Omega)= 
\langle \hat G^{\omega_+}_{{\vec{q}}_+{\vec{q}}_+^{\,\prime}} 
        \hat G^{\omega_-\, *}_{{\vec{q}}_-^{\,\prime} {\vec{q}}_-}  
\rangle$. Here we introduced \cite{Lubatsch05}  the   
center-of-mass and the relative frequencies ($\Omega$, $\omega$)  and momenta
($\vec{Q}$, $\vec{q}$) with $\omega_{1,2}\,=\,\omega\pm\Omega/2$  and $\vec
q_{1,2}\,=\,\vec q \pm\vec Q/2$. The variables $\Omega$, $\vec Q$ are
associated with the time and the position dependence of the averaged energy density, with
$\hat Q =\vec Q/|\vec Q|$, while  
$\omega_{\pm} = \omega \pm \Omega /2$ and  
$\vec{q}_{\pm} = \vec{q} \pm \vec{Q}/2$ etc. are the frequencies  
and momenta of in- and out-going waves, respectively. 
The intensity correlation, the disorder averaged 
particle-hole Green's function,
$\Phi^{\omega}_{\vec{q}\vec{q}^{\prime}}(\vec{Q},\Omega)$
is described by  the Bethe-Salpeter equation 
\begin{eqnarray}
\label{bethe_eq}
\!\!\!\!\!\!\!\!\!\!\!\!\!\!\!\!\!\!\!\!\!\!\!\!\!\!\!\!\!\!\!\!\!\!\!\!\Phi_{\vec{q}\vec{q}'}\, 
=\,
G^R_{q_+}(\omega_+)G^A_{q_-}(\omega_-) {\color{white}0000000000000000}\nonumber\\
\!
\qquad{\color{white}00000000}\left[ 
\delta(\vec q - \vec q\prime)  
+ 
\left\lmoustache     \frac { {\rm d}^3q\,''\,}    { (2\pi)^3 } \right. 
\gamma_{q\,q\,''\,}\Phi_{\vec{q}\,''\,\vec{q}\,'\,}
\right]\,.
\end{eqnarray}
By utilizing the averaged single particle Green's function, {\it c.f.} Eq. (\ref{SP_Green_function}), on the left-hand side
of Eq. (\ref{bethe_eq}) the Bethe-Salpeter equation may be rewritten as  the kinetic equation, see, e.g., 
ref.\ \cite{Lubatsch05}, 
\begin{eqnarray} 
\label{boltzmann} 
\left[ 
\omega\Omega\frac{{\rm Re}{\epsilon_b}}{c^2} 
- Q\, (\vec{q}\cdot\hat{Q}) +\frac{i}{c^2\tau ^2} 
\right] 
\Phi^{\omega}_{\vec{q}\vec{q}^{\prime}} 
&=& \nonumber\\ 
&&\hspace*{-4.7cm}- i {\rm Im} G^{\omega}_{\vec{q}} 
\left[ \delta(\vec q -\vec q\prime) + 
\left\lmoustache  \frac{{\rm d}^3 q^{\prime\prime}}{(2 \pi)^3} \right. 
 \gamma^{\omega}_{\vec{q}{\vec{q}^{\prime\prime}}} 
\Phi^{\omega}_{\vec{q}^{\prime\prime}\vec{q}^{\prime}} 
\right]. 
\end{eqnarray} 
When we analyze the correlation function's long-time 
($\Omega \to 0$) and long-distance ($ \vert \vec{Q} \vert \to 0$)  
behavior, terms of $O(\Omega^2, Q^3, \Omega Q)$ are  
neglected. Eq. (\ref{boltzmann}) contains the {\it total}  
quadratic momentum relaxation rate $1/\tau^{2}=c^2\,{\rm Im}  ( \epsilon_b  
\omega^2/c^2-\Sigma ^{\omega})$ due to absorption/gain and due to impurity
scattering in the background medium as well as the irreducible two-particle vertex function $\gamma^{\omega}_{\vec{q} \vec{q}^{\,\prime}}(\vec{Q},\Omega)$.
In order to solve this equation we expand it into moments.

The energy conservation is implemented in the solution 
of the Bethe-Salpeter equation 
by a Ward identity (WI) for the photonic case, see Ref. \cite{Lubatsch05}. The Ward identity is derived in the generalized form for
  the scattering of photons in {\it non-conserving} media.  Non-linear
  effects, absorption and gain
  yield an additional contribution, and a form of the Ward-Takahashi identity
  for photons in complex matter \cite{Lubatsch05,WARD,TAKAHASHI} is
  derived. The additional contribution is not negligible and thus effectively present in all results of the transport characteristics of the self-consistent
framework \cite{Lubatsch05,FrankPRB2006,Frank2011}. For scalar waves the Ward identity assumes
the following exact form 
\begin{eqnarray} 
\label{Ward} 
\Sigma^{\omega_+}_{\vec{q}_+} - \Sigma^{\omega_-\, *}_{\vec{q}_-} 
 \!\!&-& \!\! 
\left\lmoustache  \!\!\frac {{\rm d}^3 q^{\prime}}{(2\pi)^3} \right. 
\left[G^{\omega_+}_{{\vec{q}}_+^{\,\prime}} - G^{\omega_-\, 
  *}_{{\vec{q}}_-^{\,\prime}} \right] \,  
{ \gamma}^{\omega}_{{\vec{q}}^{\,\prime}{{\vec{q}}}}({\vec{Q}},\Omega) 
\\ 
 \! \!\!&=& \!\!\! 
f_{\omega}(\Omega)\! 
\left[\! 
{\rm Re}{\Sigma}^{\omega}_{{\vec{q}}} 
 \!+\!\! 
\left\lmoustache  \!\!\frac {\rm d^3 q^{\prime}}{(2\pi)^3} \right. 
{\rm Re} {G}^{\omega}_{{\vec{q}}^{\,\prime}} \,  
{ \gamma}^{\omega}_{{\vec{q}}^{\,\prime}{{\vec{q}}}}({\vec{Q}},\Omega) 
\right]\! . 
\nonumber 
\end{eqnarray} 
The right-hand side of Eq. (\ref{Ward}) represents reactive effects 
(real parts), originating from the explicit $\omega^2$-dependence of the 
photonic random potential. In conserving media (${\rm Im} \epsilon _b = {\rm Im} \epsilon_{scat}  =0$) these terms renormalize the energy transport velocity $v_{\mbox{\tiny
    E}}$ relative to the average phase velocity  $c_p$  without emphasizing the diffusive long-time behavior.\cite{Kroha,Lubatsch05} 
In presence of loss or gain, additional effects are enhanced by the prefactor
\begin{eqnarray} 
\label{prefactorF} 
f_{\omega}(\Omega)= \frac{
(\omega\Omega {\rm Re} \Delta\epsilon + i\omega^2 {\rm Im} \Delta\epsilon )}{
(\omega^2     {\rm Re} \Delta\epsilon + i\omega\Omega {\rm Im} \Delta\epsilon)},
 \end{eqnarray}
which now does not vanish in the limit of $\Omega \to 0$. 

\subsection{Expansion of the two-particle Green's function into moments}

The solution of the Bethe-Salpeter equation is derived by rewriting it in the
form of a kinetic equation and by deriving a continuity equation. For this aim
we expand the intensity correlator into its moments and we extract a diffusion pole  structure from the Bethe-Salpeter equation Eq. (\ref{bethe_eq}).
The $\vec{q}\,'\,$ integrated correlator
\begin{eqnarray}
\label{Phi_q}
\Phi_{\vec{q}}
=
\left\lmoustache   \! \!  \frac { {\rm d}^3q\,'\,}    { (2\pi)^3 } \right.\! \!  
\Phi_{\vec{q\,}\vec{q}\,'\,} 
\end{eqnarray}
is decoupled from the momentum dependent prefactors by some auxiliary
approximation scheme. This approximation must obey the results of the ladder
approximation and it must incorporate the set of physical relevant variables
for the observed phenomena. In a first step we use the bare first two moments
of the correlation function $\Phi_{\vec{q}}$ defined as
\begin{eqnarray}
\label{density_density}
\Phi_{\rho\rho}(\vec{Q},\Omega)
&=&
\left\lmoustache   \!  \frac { {\rm d}^3q}    { (2\pi)^3 } \right.
\!\!
\left\lmoustache   \!  \frac { {\rm d}^3q\,'} { (2\pi)^3 } \right.
\Phi_{\vec{q}\,\vec{q}\,'\,}
\\
\label{density_current}
\Phi_{j\rho}(\vec{Q},\Omega)
&=&
\left\lmoustache   \! \frac { {\rm d}^3q}  { (2\pi)^3 } \right.
\!\!
\left\lmoustache   \!\frac { {\rm d}^3q\,'}{ (2\pi)^3 } \right.
(\vec{q}\cdot\hat{Q})
\Phi_{\vec{q}\,\vec{q}\,'\,}.
\end{eqnarray} 
These bare moments are related  to  physical quantities,
the energy density correlation $P^{\omega}_{\mbox{\tiny E}}(\vec{Q},\Omega)$
and the current density correlation $J^{\omega}_{\mbox{\tiny E}}(\vec{Q},\Omega)$,
by dimensional prefactors:
\begin{eqnarray}
\label{density_density_physical}
P^{\omega}_{\mbox{\tiny E}}(\! \vec{Q},\Omega )\!\! &=&\! \! 
\left[\!
\frac{\omega}{c_{\mbox{\tiny p}}}
\!\right]^2\!\!\!  \Phi_{\rho\rho}
\,\,\,\,
\Leftrightarrow
\Phi_{\rho\rho}
\!=\! \left[\!
\frac{c_{\mbox{\tiny p}}}{\omega}
\!\right]^2\!\!P^{\omega}_{\mbox{\tiny E}}(\! \vec{Q},\Omega )
\\
\label{density_cuurent_physical}
J^{\omega}_{\mbox{\tiny E}}(\! \vec{Q},\Omega )\!\! &=& \!\! 
\left[\!
\frac{\omega v_{\mbox{\tiny E} }} {c_{\mbox{\tiny p}}}
\!\right]\!
 \Phi_{j\rho}
\Leftrightarrow
\Phi_ {j\rho}
\! =\! 
\left[\! 
\frac {c_{\mbox{\tiny p}}} {\omega v_{\mbox{\tiny E} }} 
\! \right]\!
J^{\omega}_{\mbox{\tiny E}}(\! \vec{Q},\Omega )\, .
\end{eqnarray}
The projection of the correlator $\Phi_{\vec{q}}$, Eq. (\ref{Phi_q}), onto the
bare moments $\Phi_ {\rho\rho}(\vec{q},\Omega)$ as in Eq. (\ref{density_density}),
and  $\Phi_ {j\rho}(\vec{q},\Omega)$, as in  Eq. (\ref{density_current}),
is therefore then by
\begin{eqnarray}
\label{intensity_projection}
\left\lmoustache   \! \frac { {\rm d}^3q\,'}{ (2\pi)^3 } \Phi_{\vec{q}\vec{q}\,'}  \right.
&\!\!=\!\!&
\frac{A(\vec{q}\,)}
     {
       \! \left\lmoustache   \! \!  \frac { {\rm d}^3q'}{ (2\pi)^3 }  \right. \! \! A(\vec{q}\,')}
\Phi_ {\rho\rho}(\vec{Q},\Omega)
\\
&\!\!+\!\!&
\frac{B(\vec{q}\,)  (\vec{q}\cdot\hat{Q}) }
     {
        \! \left\lmoustache   \! \!  \frac { {\rm d}^3q'}{ (2\pi)^3 } \right. \! \! 
	B(\vec{q}\,')(\vec{q}\,'\cdot\hat{Q})^2}
\Phi_ {j\rho}(\vec{Q},\Omega).\nonumber
\end{eqnarray}
The projection coefficients $A(\vec{q}\,)$ and $B(\vec{q}\,)$  are to be
determined in the following. In this expansion the bare moments may be
substituted by their physical counterparts, the energy density $P^{\omega}_{\mbox{\tiny E}}$ in Eq. (\ref{density_density_physical}) and 
the current density $J^{\omega}_{\mbox{\tiny E}}(\vec{Q},\Omega)$ from Eq. (\ref{density_cuurent_physical}).
The expansion coefficients $A(\vec{q}\,)$ and $B(\vec{q}\,)$ in
Eq. (\ref{intensity_projection}) behave uncritically when the system localizes. Thus they can be determined
by using the simple ladder approximation, where all expressions are known
exactly. The ladder approximation of the two-particle vertex function is
schematically explained in \cite{Lubatsch05,FrankPRB2006,Frank2011}.

In the following we use the approximation to obtain the expansion coefficients from it.
In the ladder approximation the zeroth bare moment is given by
\begin{eqnarray}
\Phi_ {\rho\rho}^L (\vec{Q},\Omega)
&=&
\left\lmoustache   \!  \frac { {\rm d}^3q}{ (2\pi)^3 }   \right.\! \! 
\left[ 
G_{\vec{q}_+}(\vec{Q},\Omega)G^*_{\vec{q}_-}(\vec{Q},\Omega)
\right]
^2 \Gamma_L\nonumber\\
\label{density_ladder}
&=&
\frac 1{\tilde\gamma_0^2}\Gamma_L,
\end{eqnarray}
the superscript $L$ refers to the ladder approximation. The product 
$
\left\lmoustache   \!  
\frac { {\rm d}^3q}
           { (2\pi)^3 } \right.\! \! 
\left[ 
G_{\vec{q}_+}(\vec{Q},\Omega)G^*_{\vec{q}_-}(\vec{Q},\Omega)
\right]
^2 $
has been expanded up to linear order in $\vec{q}$. 
The renormalized vertex $\tilde\gamma_0$ is given by
\begin{eqnarray}
\tilde\gamma_0
&\!\!=\!\!&
\gamma_0
\!+\!
f_{\omega}(\Omega)
\frac
{
\left(\!
{\rm Re\,}\gamma_0  G_0
\!+\!
{\rm Re\,} \Sigma \!
\right)
}
{
{\rm Im\,} G_0
}
\!- \!
\frac
{
\omega^2 {\rm Im\,} \epsilon_b
}
{
 {\rm Im\,} G_0
}
\end{eqnarray}
where $\gamma_0$ is the bare vertex. $f_{\omega}(\Omega)$ is arising from the
Ward identity and has been defined in Eq. (\ref{Ward}).
Within the simple ladder approximation the bare  moment $\Phi^L_{j\rho}(\vec{Q},\Omega)$
defined in Eq. (\ref{density_current}) is thus given by 
\begin{eqnarray}
\Phi^L_{j\rho}(\!\vec{Q},\Omega)
\!\!=\!\!
\! \left\lmoustache   \!  \frac { {\rm d}^3q}   { (2\pi)^3 }\right.\! \! 
(\vec{q}\cdot\hat{Q})
G_{\vec{q}_+}\!G^*_{\vec{q}_-}
\! \! \left\lmoustache   \!   \frac { {\rm d}^3q\,'\,} { (2\pi)^3 } \right.\! \! 
G_{\vec{q}\,'_+}\!G^*_{\vec{q}\,'_-}\!\!
\Gamma_L\,.
\end{eqnarray}
We follow this strategy again and by expanding the product $G_{\vec{q}\,'_+}G^*_{\vec{q}\,'_-}$
under the second integral up to first order in $\vec{q\,}'$
we obtain the expression
\begin{eqnarray}
\Phi^L_ {j\rho}(\vec{Q},\Omega)
=
\frac 1{\tilde\gamma_0}
\Gamma_L
\! \left\lmoustache   \!  \frac { {\rm d}^3q} { (2\pi)^3 }\right.\! \! 
(\vec{q}\cdot\hat{Q})
G_{\vec{q}_+}G^*_{\vec{q}_-}
\end{eqnarray}
By employing the same expansion to the remaining product of the Green's function one eventually finds
\begin{eqnarray}
\label{current_ladder}
\Phi^L_ {j\rho}(\vec{Q},\Omega)
=
\frac {\Gamma_L} {\tilde\gamma_0}
\! \left\lmoustache   \!  \frac { {\rm d}^3q}  { (2\pi)^3 }\right.\! \! 
(\vec{q}\cdot\hat{Q})
\frac 12
\frac 
{\Delta G_{\vec{q}}^2  (\vec{q}\cdot\hat{Q}) Q  }
{\tilde\gamma_0 \Delta G_0},
\end{eqnarray}
where the abbreviation $\Delta G \equiv G - G^*$ has been introduced.

In the next step of determining the expansion coefficients $A(\vec{q}\,)$ 
and $B(\vec{q}\,)$, Eq. (\ref{intensity_projection}), we go back to the field-field correlation function
$
\Phi_{\vec{q}\vec{q}\,'\,}$.
In the uncritical ladder approximation the two particle Green's function
is given by
\begin{eqnarray}
\label{some_phi_ladder}
\! \left\lmoustache   \!  \frac { {\rm d}^3q\,'}{ (2\pi)^3 } \right.\! \! 
\Phi_{\vec{q}\vec{q}\,'}
=
\left[
G_{\vec{q}_+}G^*_{\vec{q}_-}
\right]
\Gamma_L
\! \left\lmoustache   \!  \frac { {\rm d}^3q\,'}  { (2\pi)^3 }\right.\! \! 
G_{\vec{q}\,'_+}G^*_{\vec{q}\,'_-}.
\end{eqnarray}
Employing the momentum expansion again, the equation, Eq. (\ref{some_phi_ladder}) can be simplified to yield
\begin{eqnarray}
\label{intensity_ladder}
\Phi_{\vec{q}}
=
\frac
{\Delta G_{\vec{q}}}
{\tilde\gamma_0^2 \Delta G_0}\Gamma_L
+
\frac 12
\frac
{\Delta G_{\vec{q}}^2 (\vec{q}\cdot\hat{Q}) Q}
{\tilde\gamma_0^2 \Delta G_0}\Gamma_L.
\end{eqnarray}
By using the above given momentum expansion,
Eq. (\ref{intensity_ladder}), in combination with the expressions given 
in Eq. (\ref{current_ladder}) and in Eq. (\ref{density_ladder})
in connection with the described projection, or expansion into moments,
Eq. (\ref{intensity_projection}),  the following relation is eventually derived
\begin{eqnarray}
\label{comp_of_coeff}
&&
\frac
{\Delta G_{\vec{q}}}
{\tilde\gamma_0^2 \Delta G_0}\Gamma_L
+
\frac 12
\frac
{\Delta G_{\vec{q}}^2 (\vec{q}\cdot\hat{Q}) Q}
{\tilde\gamma_0^2 \Delta G_0}\Gamma_L
\\\nonumber
&&=\!
\frac{A(\vec{q}\,)}
     {
       \! \left\lmoustache   \! \! \frac { {\rm d}^3q'}  { (2\pi)^3 }  \right.\! \!   A(\vec{q}\,')}
\frac 1{\tilde\gamma_0^2}\Gamma_L
\\
&&+
\frac{B(\vec{q}\,)  (\vec{q}\cdot\hat{Q}) }
     {
       \! \left\lmoustache   \! \!\frac { {\rm d}^3q'}{ (2\pi)^3 }  \right.\! \!  
       B(\vec{q}\,')(\vec{q}\,'\cdot\hat{Q})^2}
\frac {\Gamma_L}{\tilde\gamma_0}
 \! \left\lmoustache   \! \frac { {\rm d}^3q} { (2\pi)^3 } \right.\! \!  
(\vec{q}\cdot\hat{Q})
\frac 12
\frac 
{\Delta G_{\vec{q}}^2  (\vec{q}\cdot\hat{Q}) Q  }
{\tilde\gamma_0 \Delta G_0}\,.
\nonumber
\end{eqnarray}
By comparison of coefficients in the relation, Eq. (\ref{comp_of_coeff}),
the  demanded coefficients $A({\vec{q}}\,)$ and $B({\vec{q}}\,)$ of the 
expansion into moments, Eq. (\ref{intensity_projection}), can now be determined 
as follows
\begin{eqnarray}
A({\vec{q}}\,)
=\Delta G_{\vec{q}}
\qquad
B({\vec{q}}\,)
=\Delta G_{\vec{q}}^2.
\end{eqnarray}
%
%
Employing those expressions for the expansion coefficients, we can eventually express
the two-particle correlator $\Phi_{\vec{q}\vec{q}\,'}$ as follows
\begin{eqnarray}
\label{decoupled_projection_physical}
\! \left\lmoustache   \!  \frac { {\rm d}^3q'}{ (2\pi)^3 }  \right.\! \!  
\Phi_{\vec{q}\vec{q}\,'}
&=&
\frac{
         \Delta G_{\vec{q}}
     }
     {
       \left(
       \frac{\omega}{c_{\mbox{\tiny p}}}
       \right)^2 
       \! \left\lmoustache   \! \! 
       \frac { {\rm d}^3q\,'} { (2\pi)^3 } \right.\! \!  
       \Delta G_{\vec{q}\,'}   
     }
P^{\omega}_{\mbox{\tiny E}}(\vec{Q},\Omega)
\\
&&+
\frac {
         \Delta G_{\vec{q}}^2 (\vec{q}\cdot\hat{Q}) 
      }
      {
	\left(
	\frac{\omega v_{\mbox{\tiny E} }} {c_{\mbox{\tiny p}}}
	\right)
	\! \left\lmoustache   \! \! 
	\frac { {\rm d}^3q\,'}{ (2\pi)^3 } \right.\! \!  
	\Delta G_{\vec{q}\,'}^2 (\vec{q}\,'\cdot\hat{Q})^2
      }
J^{\omega}_{\mbox{\tiny E}}(\vec{Q},\Omega).
\nonumber
\end{eqnarray}
The expression, Eq. (\ref{decoupled_projection_physical}), 
represents the complete expansion of the intensity correlator into 
its moments.
This form is used now to decouple and therefore solve the Bethe-Salpeter equation.


\subsection{General solution of the Bethe-Salpeter equation}

We repeat the most important steps so far. The disorder averaged intensity correlation, the two-particle Green's function,
obeys the Bethe-Salpeter equation, see Eq. (\ref{bethe_eq})
\begin{eqnarray}
\Phi_{\vec{q}\,\vec{q}\,'\,} 
\!\!&=&\!\! 
G_{q_+}^{\omega_+}G^{*\,\omega_-}_{q_-}
\!
\left[ 
1 \! \!  
+ \! \!  \!  
\left\lmoustache   \! \!  \frac { {\rm d}^3q\,''\,}    { (2\pi)^3 } \right.\! \!  
\gamma_{q\,q\,''\,}\Phi_{\vec{q}\,''\,\vec{q}\,'}
\right]
\end{eqnarray}
%
The Bethe-Salpeter equation may be rewritten
into the kinetic or Boltzmann equation given in Eq. (\ref{boltzmann})
\begin{eqnarray}
\label{kinetic}
&&\!\!\!\!\!\!\!\!\!\!\!\!\!\!\!\!\!\!\!
\left[ (\omega\Omega)\, 2 {\rm Re\,}\epsilon- Q \left(\vec{q}\cdot\hat{Q}\right) + \Delta \Sigma -\omega^2 \Delta\epsilon  \right] 
\Phi_{\vec{q}}
\nonumber\\
&&\qquad
=
\Delta G_{\vec{q}} + 
\left\lmoustache   \! \!  \frac { {\rm d}^3q\,'}    { (2\pi)^3 } \right.\! \!  
\Delta G_{\vec{q}} \gamma_{\vec{q}\vec{q}\,'\,}\Phi_{\vec{q}\,'\,}\,.
\end{eqnarray}
To find the solution of Eq. (\ref{kinetic}), we first sum in
Eq. (\ref{kinetic}) over momenta $\vec{q}$, and we incorporate the generalized  
Ward identity as given in Eq. (\ref{Ward}) and we expand the obtained result 
for small internal momenta $Q$ and internal frequencies $\Omega$.
It is essential to employ the form of the two-particle correlator shown in
Eq. (\ref{decoupled_projection_physical}). Eventually the generalized
continuity equation for the energy density can be derived as
%
%
%
\begin{eqnarray} 
\Omega P^{\omega}_{\mbox{\tiny E}} + Q J^{\omega}_{\mbox{\tiny E}} =  
\frac{4\pi i \,\omega\, N(\omega )} 
       {g^{(1)}_{\omega}\left[ 1 + \Delta(\omega) \right] c_p^{2}} 
\!\!&+&\!\! \frac{i [g^{(0)}_{\omega} + \Lambda(\omega) ]} 
       {g^{(1)}_{\omega}\left[ 1 + \Delta(\omega) \right] } 
       P^{\omega}_{\mbox{\tiny E}}.
\nonumber\\  
\label{continuityL}
\end{eqnarray}
The generalized continuity equation represents energy conservation in the
presence of optical gain and/or absorption.

As the standard solution procedure the next step is to obtain
a linearly independent equation which relates the energy density
$P^{\omega}_{\mbox{\tiny E}}$ and the current density $J^{\omega}_{\mbox{\tiny
    E}}$. This is realized in a similar way to above, one first multiplies the kinetic
equation, Eq.(\ref{kinetic}), by the projector $\left[\vec{q}\cdot\hat{Q}\right]$ and then follows the already outlined steps to eventually obtain
the wanted second relation. This is the current relaxation equation
\begin{eqnarray} 
\label{CDR} 
\left[\omega\Omega\frac{{\rm Re}{\epsilon_b}}{c^2} 
      +\frac{i}{c^2\tau ^2}+iM(\Omega) 
\right] 
J^{\omega}_{\mbox{\tiny E}} \!\!&+&\!\! 
\tilde A\, Q P_{\mbox{\tiny E}}^{\omega} =0\  
,
\end{eqnarray} 
relating the energy density $P^{\omega}_{\mbox{\tiny E}}$ 
and the energy density current $J^{\omega}_{\mbox{\tiny E}}$. The memory function $M(\Omega)$ is introduced according to
\begin{eqnarray} 
\label{M_Omega}
M\!(\Omega ) 
\!=\!
\frac
{i\!\!
\left\lmoustache  \!\!\frac {{\rm d}^3 q}{(2\pi)^3} \right.
\!\!\!
\left\lmoustache  \!\!\frac {{\rm d}^3 q\,'}{(2\pi)^3} \right. 
\!\!
[\vec{q}\!\cdot\!\hat{Q}] 
\Delta G_{\vec{q}}^{\omega}
\gamma^{\omega}_{\vec{q}\vec{q}\,'}
(\Delta G_{\vec{q}\,'}^{\omega})^2
[\vec{q}\,'\!\cdot\!\hat{Q}] 
}
{
\left\lmoustache  \!\!\frac {{\rm d}^3 q}{(2\pi)^3} \right. 
[\vec{q}\!\cdot\!\hat{Q}]^2 (\Delta G_{\vec{q}}^{\omega})^2 
},
\end{eqnarray} 
where $\gamma^{\omega}_{\vec{p}\vec{p}'}\equiv \gamma^{\omega}_{\vec{p}\vec{p}'}(\vec{Q},\Omega)$
is the total irreducible two-particle vertex, which will be discussed in
detail in what follows.

So far, two independent equations, Eq. (\ref{continuityL}) and Eq. (\ref{CDR}),
have been obtained. Either of them is relating 
the current density $J^{\omega}_{\mbox{\tiny E}}$ with the energy density
$P^{\omega}_{\mbox{\tiny E}}$. Now one can eliminate one of the two variables in this
linear system of equations. The two equation are combined to find an expression for the energy density 
\begin{eqnarray} 
\label{P_E} 
P_{\mbox{\tiny E}}^{\omega}(Q,\Omega) = 
\frac{4\pi i N(\omega) /  
(g^{(1)}_{\omega}\left[ 1 + \Delta(\omega) \right] c_p^{2})} 
{\Omega + i Q^2 D + i \xi_a^{-2}D}\ , 
\end{eqnarray} 
that exhibits  the expected diffusion pole structure for
non-conserving media. Precisely in the denominator of Eq. (\ref{P_E})
there appears an additional term as compared to the case of conserving media.
This is the term $ \xi_a^{-2}D$, the mass term,  accounting for loss 
(or gain) to the intensity not being due to diffusive relaxation.
In  Eq. (\ref{P_E}) also  the generalized, 
$\Omega$-dependent diffusion coefficient $D(\Omega)$ has been
introduced by the relation 
\begin{eqnarray} 
\label{D_omega_full} 
D(\Omega)  
\left[1 - i \, \Omega \omega \tau ^2 {\rm Re}\epsilon_b  \right] =   
D_0^{tot} -c^2\tau^2 D(\Omega ) M(\omega ).
\end{eqnarray} 
It shall be noted that  Eq. (\ref{P_E}) introduces
the absorption or gain induced growth or absorption
scale $\xi_a$ of the diffusive modes, 
\begin{eqnarray}  
\label{xi_a} 
\xi_a^{-2}&=& 
\frac
{r_{\epsilon} A_{\epsilon} -2\omega^2{\rm Im} \epsilon_b} 
{2{\rm Re}\epsilon_b-A_{\epsilon}B_{\epsilon}/\omega}\ 
\frac {1} {\omega D(\Omega)},
\end{eqnarray} 
which is to be well distinguished from the single-particle or amplitude
absorption or amplification length.
The diffusion constant without memory effects in Eq. (\ref{D_omega_full}),  
$D_0^{tot}=D_0+D_b+D_{scat}$, consists of the bare diffusion constant \cite{Kroha}, 
\begin{eqnarray} 
D_0 = \frac  
{2v_{\mbox{\tiny E}} c_p} 
{ \pi N(\omega)} 
\left\lmoustache  \!\!\frac {{\rm d}^3 q}{(2\pi)^3} \right. 
[\vec{q}\cdot\hat{Q}]^2 ({\rm Im} G_{\vec{q}}^{\omega})^2 
\label{Dbare} 
\end{eqnarray} 
and renormalizations from absorption or gain in the background medium 
($D_b$) and in the scatterers ($D_{scat}$), 
\begin{eqnarray} 
\label{D_B} 
D_b =  \left(\omega\tau \right)^2  \,{\rm Im} \epsilon_b\, \tilde{D}_0/4 \ ,
\quad 
D_{scat} = r_{\epsilon}A_{\epsilon}\tau^2 \tilde{D}_0/8 \ , 
\end{eqnarray} 
where $\tilde D_0$ is the same as in Eq. (\ref{Dbare}), 
with $\Delta
G(Q,\Omega)=(G_{\vec{q}}^{\omega})^A-(G_{\vec{q}}^{\omega})^R$
replaced by $\Box
G(Q,\Omega)=(G_{\vec{q}}^{\omega})^A+(G_{\vec{q}}^{\omega})^R$
\cite{FrankPRB2006,Frank2011,ApplSci}, and $({\rm Im} G_{\vec{q}}^{\omega})^2$ replaced by 
${(\rm Re} G_{\vec{q}}^{\omega})^2$, respectively. In the above Eqs. (\ref{xi_a})-(\ref{D_B}) 
the following short-hand notations have 
been introduced, 
%
\begin{eqnarray}
u _{\epsilon}&=&\frac {{\rm Im} (\Delta\epsilon \Sigma^{\omega})}
                 {{\rm Im} (\Delta\epsilon G_0^{\omega}) }
\ ,\qquad\qquad
r_{\epsilon} = {{\rm Im} \Delta\epsilon}/{{\rm Re} \Delta\epsilon},
\nonumber\\
A_{\epsilon} &=& 2 [u _{\epsilon} {\rm Re} G_o +  {\rm Re} \Sigma_o] \nonumber\\
B_{\epsilon} &=& \frac{({\rm Re}\Delta\epsilon)^2+({\rm Im}\Delta\epsilon)^2}
{2\omega^2({\rm Re}\Delta\epsilon)^2}.
\nonumber
\end{eqnarray}


\subsection{Vertex function and self-consistency}
\label{VERTDIS}

\begin{figure}[t] 
\begin{center} 
\includegraphics[width=0.75\linewidth]{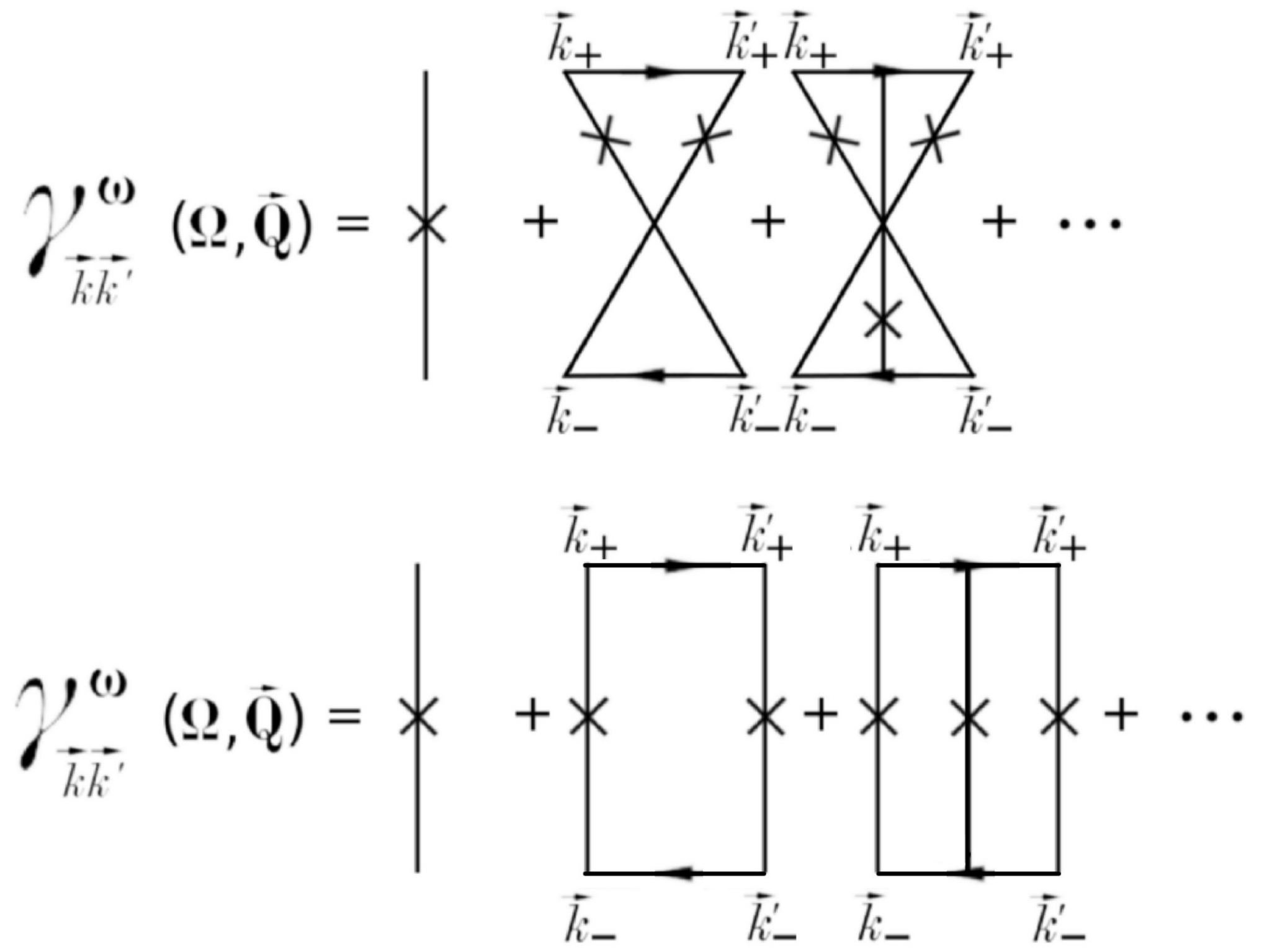} 
\end{center} 
\vspace*{-0.4cm}
\caption{
         The upper panel shows a diagrammatic expansion
	 of the irreducible two-particle vertex $\gamma$. The bottom panel 
	 displays the disentangled Cooperon with changed momentum arguments.}
\label{Cooperon_Flip} 
\end{figure} 

According to Eq. (\ref{M_Omega}) and Eq. (\ref{D_omega_full}) 
the energy density, the two-particle function, given in Eq. (\ref{P_E}) still depends
on the full two-particle vertex $\gamma^{\omega}_{{\vec{q}}^{\,\prime}{{\vec{q}}}}$.
Before discussing the vertex function, we briefly recall our arguments with
regards to dissipation. Dissipation breaks the time reversal symmetry
\cite{OnsagerI,OnsagerII} on the one hand side but on the other hand the dissipation rate
itself is invariant under time reversal. As a general picture of this physics
the damped  harmonic oscillator can be mentioned, where  the time reversed
solution is still damped with the very same damping constant. Having this in
mind we analyze the irreducible vertex  $\gamma^{\omega}_{{\vec{q}}^{\,\prime}{{\vec{q}}}}$ 
for the self-consistent calculation of $M(\Omega )$, exploiting time reversal symmetry of propagation in the active medium.
In the long-time limit ($\Omega\to 0$) the dominant contributions to 
$\gamma^{\omega}_{{\vec{q}}^{\,\prime}{{\vec{q}}}}$ are maximally crossed
diagrams (Cooperons), which are valid as well as for conserving media, and
they may be also disentangled. 

In Fig. (\ref{Cooperon_Flip}) the disentangling of the  Cooperon into 
the ladder diagram is shown. The internal momentum argument of the 
disentangled irreducible vertex function in the second line
of  Fig. (\ref{Cooperon_Flip})
is replaced by the new momentum $\vec{Q} = \vec{k} + \vec{k}'$.
Thus  $\gamma^{\omega}_{{\vec{q}}^{\,\prime}{{\vec{q}}}}$ acquires now the
absorption or gain-induced decay or growth rate $\xi_a^{-2}D$. 
Finally the memory kernel  $M(\Omega)$ reads   
\begin{eqnarray} 
M(\Omega ) &=& 
- \frac{(2v_{\tiny E}c_p)^2\ 
u_{\epsilon} \left[ 
2\pi \omega u_{\epsilon} N(\omega ) + r_{\epsilon}A_{\epsilon}  
- 2\omega^2{\rm Im}\epsilon_b 
\right] 
}
{\pi \omega N(\omega) D_0 D(\Omega)} 
\nonumber\\ 
&& 
\hspace*{-1.5cm} 
\times   
\left\lmoustache \!\frac {{\rm d}^3 q}{(2\pi)^3} \right. 
\left\lmoustache \!\frac {{\rm d}^3 q'}{(2\pi)^3}\right. 
\frac{ 
[\vec{q}\cdot\hat{Q}] 
|{\rm Im} G_q| \left({\rm Im} G_{q'}\right)^2 
[\vec{q}\,'\cdot\hat{Q}] 
} 
{\frac{-i\Omega}{D(\Omega)} + \left(\vec{q}+\vec{q}\,' \right)^2 +  
\xi_a^{-2} }\, . 
\label{MD} 
\end{eqnarray} 
Eqs.\ (\ref{D_omega_full})-(\ref{MD})  
constitute the self-consistency equations for the diffusion coefficient 
$D(\Omega )$ including the growth/decay length scale $\xi _a$ in presence of dissipation or gain.

\subsection{Length and time scales}

Within disordered systems a multitude of length an d time scales are defined,
that are related to the single or the two-particle quantities respectively. An
important length scale which can be directly measured in the experiment is the
{\it scattering mean free path} $l_s$ defined in the single particle
Green's function
\begin{eqnarray}
\label{Green_ls}
G_{\vec{q}} (\omega) 
=
\frac
{1}
{\frac{\omega^2}{c^2}\epsilon_0 -q^2 - \Sigma (\omega)  }
\end{eqnarray}
where the imaginary part of the self-energy introduces the decay length $l_s$
\begin{eqnarray}
q  &=& \frac {\omega}{c}\sqrt{\epsilon_0} \longrightarrow {\rm Re } (q) + \frac {i}  {2 l_s }    \\
l_s &=&            \frac
                  {1}
                  { 2 {\rm Im } (\sqrt{q^2+i{\rm Im } \Sigma (\omega) } )
                  }
\label{LS}
\end{eqnarray}
The decay length may equivalently be interpreted as the life time of the
corresponding k-mode. In the case where the dielectric constant is purely
real, so in the case of passive matter, the scattering mean free path $l_s$
describes the scale for determining the loss sole to scattering out of a given
k-mode. In the other case, for gain and dissipation, the k-mode experiences
amplification or absorption. In the case of gain this transport theory is
valid for ${\rm Im} \Sigma(\omega)\,<\,0 $ while the flip of ${\rm Im}
\Sigma(\omega)$, ${\rm Im}
\Sigma(\omega)\,=\,0 $, defines the point of the phase transition, i.e. the laser
threshold, for the pumped single scatterer \cite{Lagendijk_mie,FrankPRB2006,Frank2011,NJP14,SREP15,ApplSci}. 

We discuss in what follows the transport of the intensity and the scales related to it.
The two-particle Green's function as given in Eq. (\ref{P_E} ) contains two obvious scales 
originating solely  from finite values of the gain/absorption coefficient. 
These length scales may be defined by
\begin{eqnarray} 
\label{def_ell_a}
\ell_a = \frac {2\pi}{{\rm Re } (\sqrt{1/\xi_a^2}  )}\\
\ell_{osc} = \frac {2\pi}{{\rm Im } (\sqrt{1/\xi_a^2}  )}
\end{eqnarray}
where $\ell_a$ represents the amplification or absorption length of the intensity
and $\ell_{osc}$ marks the length over which the intensity oscillates,
where $\xi_a^2$ has already been defined in Eq. (\ref{xi_a} ).
The corresponding time scales may then be defined as
\begin{eqnarray} 
\label{def_tau_a}
\frac 1 {\tau_a}     &=& \frac { D} { \xi_a^2} \\
\frac 1 {\tau_{osc}} &=&       Q^2 {\rm Im } D 
\end{eqnarray}

Including the gain induced growth rate $\tau_a$ as defined in
Eq. (\ref{def_tau_a}), the intensity Green's function Eq. (\ref{P_E} ) may  now be rewritten as
\begin{eqnarray} 
\label{P_E_tau}
P(Q,\Omega) = 
\frac
{\alpha}
{-i\Omega + iQ^2{\rm Im } D + Q^2{\rm Re } D - 1/{\tau_a} }
\end{eqnarray}
where the coefficient $\alpha$ may symbolically contain all  the factors
explicitly shown and discussed in Eq. (\ref{P_E}).

Our aim in this article is to calculate the electrical field-field correlator at different positions and 
frequencies Eq. (\ref{bethe_eq}) eventually leading to the evaluation the two particle Green's function given in  Eq. (\ref{P_E}).
The momentum $Q$ appearing in  Eq. (\ref{P_E}) represents in Fourier space a
relative position within the sample. In three dimensions the momentum $Q$
actually defines a volume unit within the sample. This volume is carefully to be distinguished from all other length scales 
e.g. the sample volume etc. It is merely the scale which determines the
presence of correlation effects in photonic transport.

In analogy to the flip of the resonance, ${\rm Im } \Sigma = 0$, in the single particle Green's function
the equivalent threshold condition for the energy density is found as follows
\begin{eqnarray}
  Q^2{\rm Re } D - 1/{\tau_a} &\ge& 0  \\
\Leftrightarrow 
\frac
{4\pi^2}
{R_{crit}^2}
{\rm Re } D
- 1/{\tau_a} &=& 0.  
\end{eqnarray}
It leads to the critical length scale
\begin{eqnarray}
\label{def_R_crit}
R_{crit}
=
2\pi \sqrt{\tau_a {\rm Re } D}. 
\end{eqnarray}
This length describes the volume where photonic transport, i.e. the energy density or intensity,
may compensate diffusive losses by amplification due to the presence of some finite
optical gain.

\subsection{Weighted essentially non-oscillatory solver (WENO)}

For the time and space dependent solution of the diffusion equation,
Eq. (\ref{P_E_tau}), including a coherent laser
pulse, see Fig.(\ref{Setup}), we use a  weighted
essentially non-oscillatory method (WENO) in time in combination with a fourth
order Runge-Kutta method in space.

Like the discontinuous Galerkin method \cite{Warburton} for
hydrodynamic systems, the WENO method \cite{Shu} has been specifically
developed for discontinuous and rogue
processes, shocks and steep gradients. Such procedures are well known to cause numerical problems or
oscillations in the calculation of the first derivative. An efficient method is thus needed to
refine the discretization of the problem locally in space and time. WENO is as
such an upgrade of the essentially non-oscillatory method (ENO)
\cite{Harten,Osher} which has been
developed for the calculation of hyperbolic conservation laws. ENO replaces the 
calculation of higher-order difference quotients by the
calculation of a bunch of lower order difference quotients which are of equal
order. Whereas ENO incorporates only the difference quotient with the smallest
approximation and thus always the influence of a part of the supporting points
of a number of cells of the so-called {\it stencil} is neglected in the search
for the stencil with the smoothest result for the interpolation, the WENO
method is more sophisticated. It uses a convex combination of all candidates of lower order difference
quotients of the stencil with an attributive weight $g_i$ \cite{Liu}

\begin{equation}
g_i=\frac{1}{h + D_i^2}
\end{equation}

where $D_i$ are the smoothness indicators of the
stencil. The variable $h\,>\,0$ is defined as
the machine accuracy which prohibits a division by $0$. All weights are normalized to unity.

\begin{equation}
\tilde g_i=\frac{g_i}{\sum_jg_j}
\end{equation}

If the stencil contains a discontinuity the smoothness indicator should be
essentially 0. The convergency of the stable solution is guaranteed by the Lax equivalence theorem \cite{Lax}.

\section{Results and discussion}
\label{sec:NUMSOL}

\subsection{Scattering mean free path $l_s$ and diffusion constant $D$ for
  mono- and polydisperse passive and active scatterers}

First we discuss here as results the scattering mean free path $l_s$, see Eq. (\ref{LS}),
and the diffusion constant $D$, see Eq. (\ref{D_omega_full}), as the
self-consistent material
characteristics of the disordered sample of complex Mie scatterers
\cite{Busch,Tip1,Tip2}. As the materials initial parameters we refer in what follows to
the literature value of the passive refractive index for titania TiO$_2$,
$n\,=\,2.7$. The
    scatterers' background is air, $\epsilon_b\,=\,1$.

\begin{figure}[t]
\hspace*{0.1cm}\resizebox{0.495\textwidth}{!}{%
  \centering\includegraphics{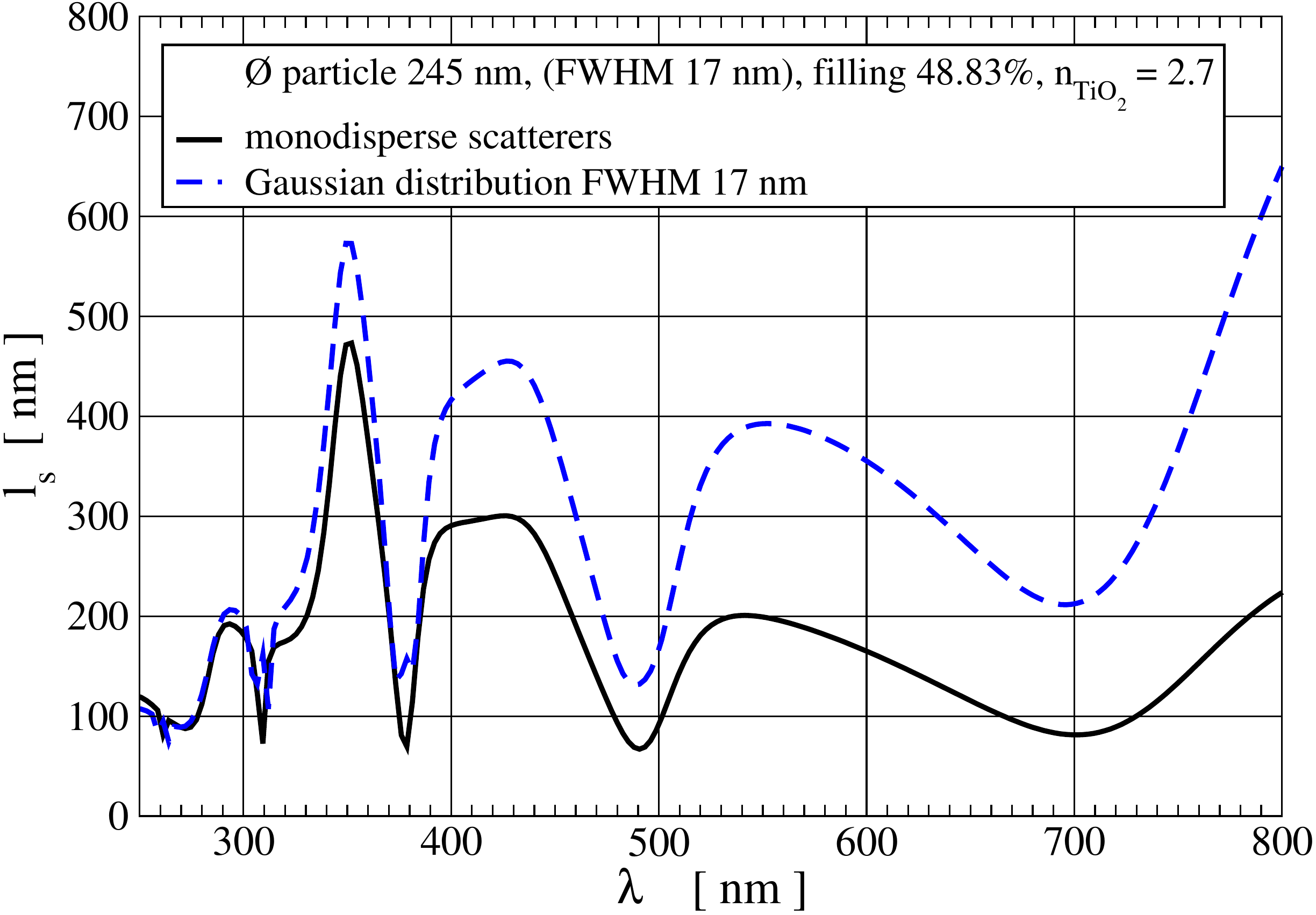}}
\vspace{0.5cm}       
\caption{Scattering mean free path $l_s\,=\,\frac{1}{2{\rm
        Im}[{\sqrt{q^2+i{\rm Im}\Sigma(\omega)}}]}$ for two disordered samples
    of TiO$_2$ Mie spheres, the passive refractive index is $n\,=\,2.7$, the
    scatterers' background is air, $\epsilon_b\,=\,1$. The blue dashed line shows the result for $l_s$ for a
    Gaussian distribution of scatterer sizes centered in the diameter $2\,r_{scat}\,=\,245.0\,nm$, full width half maximum is $17.0\,nm$, see
    Fig. (\ref{Setup}). The filling fraction is $48.83 \%$. The black line
    shows the result for $l_s$ for monodisperse Mie scatterers of the diameter
    of  $2\, r_{scat}\,=\,245.0\,nm$, all other parameters are the same. We find that the scattering mean free path $l_s$ for the Gaussian
    distribution of scatterers is overall increased compared to the
    monodisperse scatterers, the quality of the results persists,
    however the very pronounced Mie resonances, the sharp dips of $l_s$, for
    $\lambda\,=\,310.0\,nm$ and $\lambda\,=\,380.0\,nm$ are reduced.}
\label{LABEL0}       
\end{figure}

In Fig. (\ref{LABEL0}) we show the scattering mean free path $l_s$ for a Gaussion
distribution of Mie scatterers, see Fig. (\ref{Setup}), which is centered at
the radius of $r_{scat}\,=\,122.5\,nm$, the full width half maximum is
$17.0\,nm$, and we compare it to results for monodisperse
Mie scatterers of size $r_{scat}\,=\,122.5\,nm$. The description takes
advantage of the fact that scattering matrices of independent scatterers are
additive in general. We find that the principal Mie characteristics of the
central particle size $r_{scat}\,=\,122.5\,nm$ is qualitatively
conserved but quantitatively reduced. This is intuitively clear due to the additivity of scattering
matrices in the independent scatterers approach since no additional structural
effects, e.g. in the sense of a varying
concentration of surface defects or the occurrence of correlated clusters and
glass transitions, are considered so far to induce any
additional dependencies. The exact Mie resonance positions, the minima of
$l_s$, remain spectrally fixed for the Gaussian distribution of polydisperse
scatterers ensembles compared to the monodisperse ensembles. What can be definitely deduced
is that the scattering mean free path  $l_s$ overall is prolonged for the
Gaussian distribution which means
that the scattering strength of the disordered sample is effectively reduced
for polydisperse media. For wavelengths $\lambda\,=\,490.0\,nm$,
$\lambda\,=\,540.0\,nm$, and $\lambda\,=\,700.0\,nm$ a reduction of the factor of 2
and more in the magnitude of the scattering mean free path $l_s$ is
derived. The filling fraction for the result of Fig. (\ref{LABEL0}) is kept
constant at $48.83\,\%$. The scattering strength and thus also the probability to reach the regime of strong
localization of light are significantly reduced for narrow peaked
polydisperse scatterers distributions compared to monodisperse ensembles.

\begin{figure}[t]
\resizebox{0.5\textwidth}{!}{%
  \includegraphics{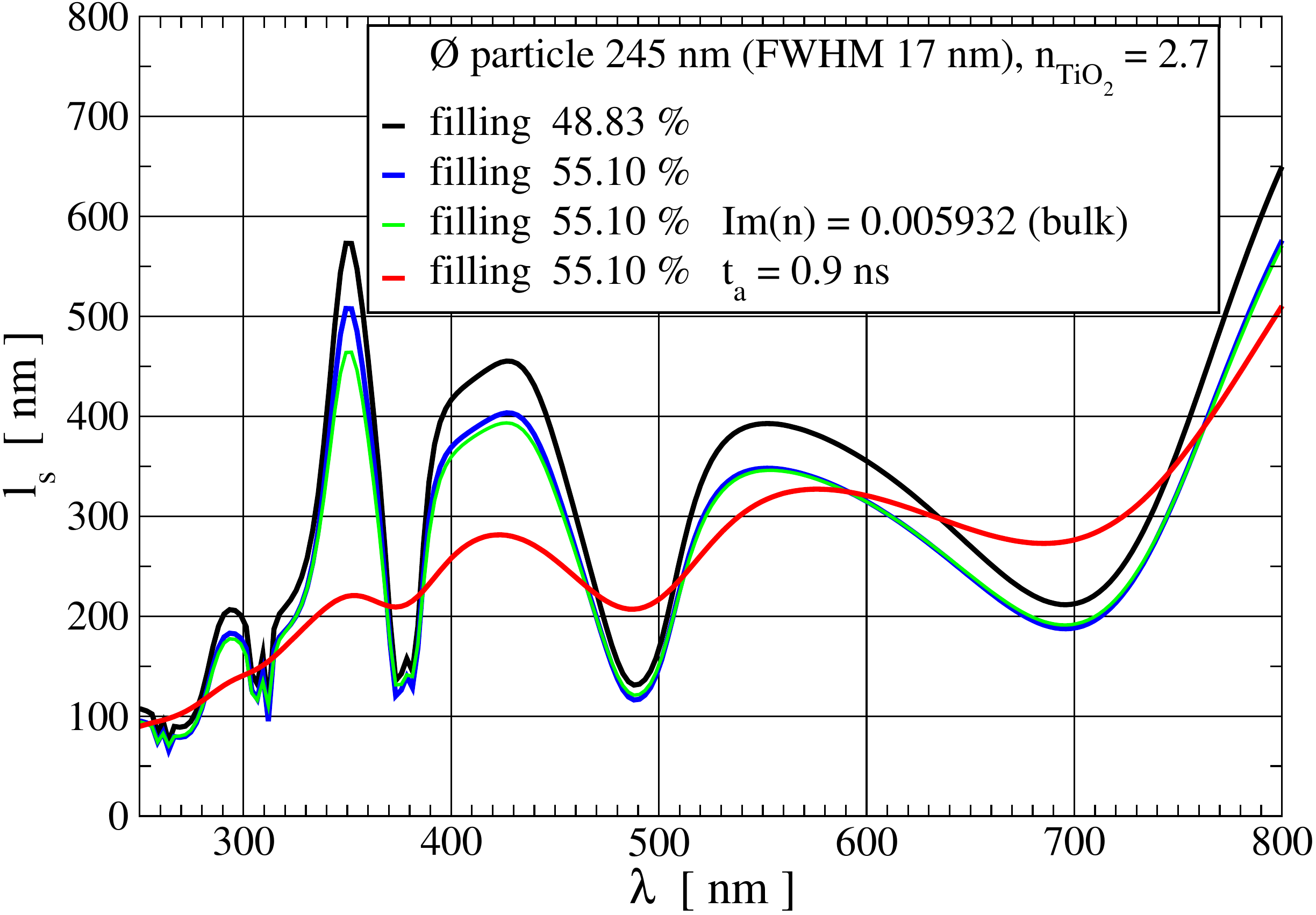}}
\vspace*{0.5cm}
\caption{Scattering mean free path $l_s$ for disordered samples of TiO$_2$ Mie
  spheres, $n\,=\,2.7$, with a Gaussian distribution of scatterers peaked at $2\,r_{scat}\,=\,245.0\,nm$, full width half maximum is $17.0\,nm$, see
    Fig. (\ref{Setup}). The
    scatterers' background is air, $\epsilon_b\,=\,1$. We display results for the filling fractions of $48.83\,
    \%$ and $55.10\, \%$ of passive scatterers and we show for the filling
    fraction of $55.10 \%$ results for active
    scatterers. The green line is the result for the literature value of
    absorption ${\rm Im}(n)\,=\,  0.005932$ for bulk, the red line is a result
    for $\tau_a\,=\,0.9\,ns$ which is equivalent to ${\rm Im}(n) =  0.3$. We
    find with an increasing filling fraction an overall decrease of $l_s$
    while its qualitative behavior is conserved. We find that the increase of
    $\tau_a$, so the increase of loss, that the Mie resonances wash out.
    However it is already visible, that all results for one specific filling
    fraction cross in what can be mathematically identified as the
    inflection points. This behavior is confirmed by the behavior of the
    diffusion coefficient $D$, see Fig.{\ref{D_GAUSS}}.}
\label{L_S_GAUSS}
\end{figure}

\begin{figure}[t]
\vspace*{0.0cm}\rotatebox{0}{\scalebox{0.36}{\includegraphics[clip]{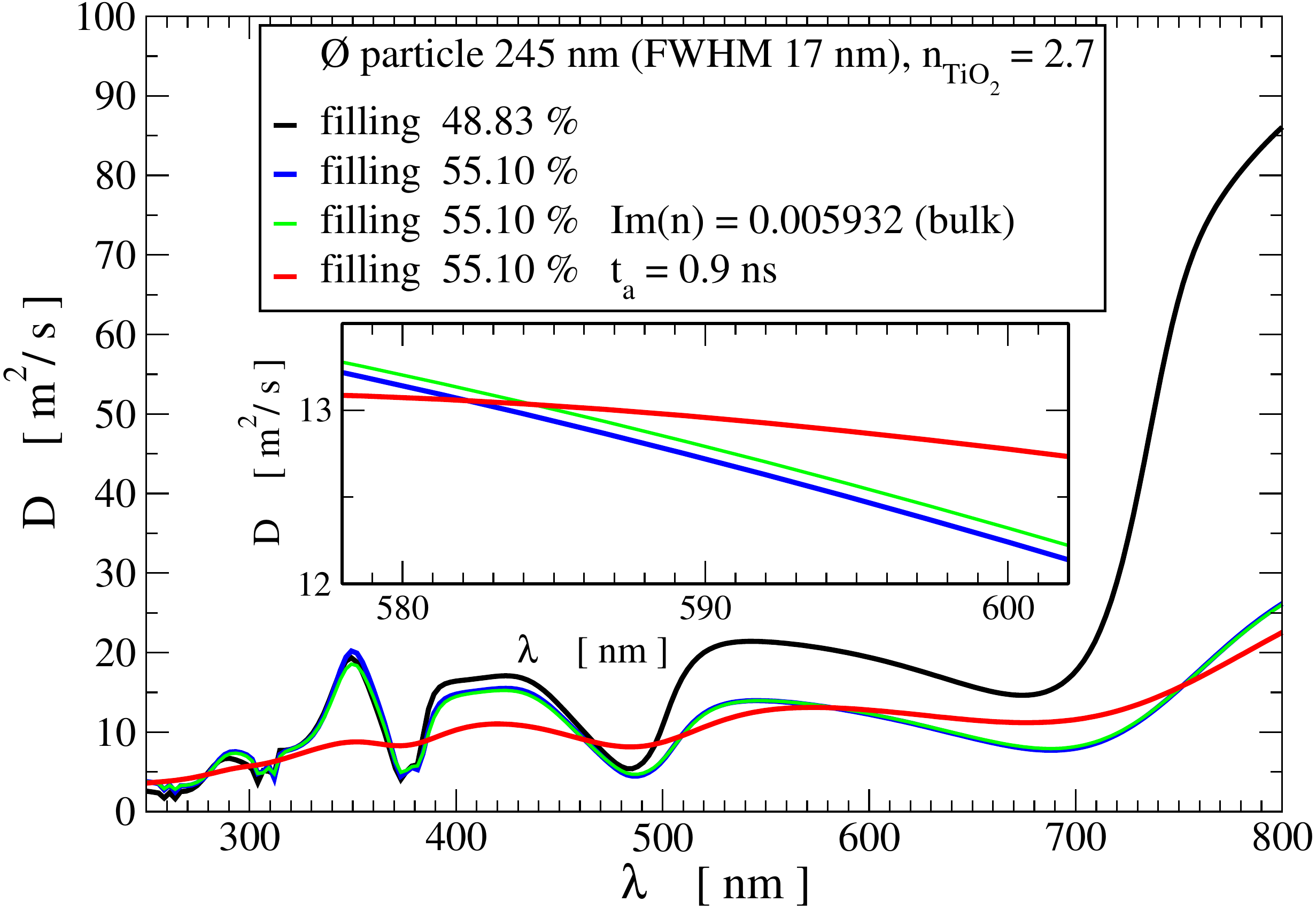}}}
\vspace*{0.0cm}
\caption{Real part of the diffusion coefficient $ReD$, see Eq. (\ref{D_omega_full}), for samples of disordered TiO$_2$ samples, parameters are equal to
  Fig.{\ref{L_S_GAUSS}}. We find that the increase of the filling fraction
  rather  moderately effects the quantitative behavior of ${\rm Re} D$ in the
  spectral region below $\lambda < 490.0\,nm$, so $\lambda\,<\,4\cdot r_{scat}$,
  for $\lambda > 490.0\,nm$ we find, that the influence of the
  filling increases. We find that the effect of the bulk value for absorption
  of ${\rm Im}(n)\,=\, 0.005932$ on the diffusion coefficient for moderate
  excitation is peaked at $\lambda\,=\,350.0\,nm$ (green line), while it is almost
  undetectable in other spectral positions. The increase of absorption to
  $\tau_a\,=\,0.9\,ns$ results in the same crossing point as found for $l_s$ at the
  inflection points. Its physical meaning is rather
  important. Absorption or gain will rather hardly
  be measured at the inflection points. In the {\it inset} we show the crossing at $\lambda\,=\,583.0\,nm$
  which has been confirmed in \cite{Maret2012,NPHOT2013} experimentally. It is
found that a value of absorption, as it is commonly detected in experiments
with TiO$_2$
powders \cite{Maret2012,NPHOT2013} can eliminate almost completely the size
dependent influences of the scatterers
geometry in the diffusion characteristics of random ensembles.}
\label{D_GAUSS}
\end{figure}

In Fig. (\ref{L_S_GAUSS}) we show the results for the scattering mean free
path $l_s$ for the Gaussian distribution of scatterers of the filling
fractions $48.83\,\%$ and $55.10\,\%$ for passive Mie resonators as well as for
absorbing scatterers. For the absorption we consider the literature value for
TiO$_2$ bulk of ${\rm Im}(n)\,=\, 0.005932$ as well as experimentally relevant
absorption values for disordered granular arrangements. 
We find that the
material characteristics of the scattering mean free path $l_s$ for an
increase of the filling fraction from  $48.83\,\%$ to $55.10\,\%$ is
quantitatively reduced, whereas is is qualitatively confirmed. No crossover
between both results is found all over the spectrum. For moderate absorption of a literature value for bulk titania, ${\rm Im}(n)\,=\, 0.005932$, the
magnitude of $l_s$ generally persists, however it is already visible for
$\lambda\,=\,260.0\,nm$ to $\lambda\,=\,270.0\,nm$, $\lambda\,=\,305.0\,nm$ to
$\lambda\,=\,320.0\,nm$, $\lambda\,=\,370.0\,nm$ to $\lambda\,=\,385.0\,nm$,
and for $\lambda\,=\,485.0\,nm$ to $\lambda\,=\,495.0\,nm$  that the Mie
resonances are washed out and $l_s$  is increasing. 
The peak positions of the characteristics at $\lambda\,=\,345.0\,nm$ to
$\lambda\,=\,360.0\,nm$, $\lambda\,=\,430.0\,nm$ to $\lambda\,=\,445.0\,nm$
show a decrease of  $l_s$ for moderate absorption. By increasing the absorption up to the value of ${\rm Im}(n)\,=\, 0.3$ which is corresponding
to $\tau_a\,=\,0.9\,ns$ we find that the Mie resonances are
further reduced. For the filling
fraction of $55.10\,\%$ there is a crossover of all results of $l_s$ found. It can be deduced that for one specific distribution
of Mie scatterers sizes and varying loss parameters and identical parameters
otherwise spectral points or at least narrow spectral regions exist where the
loss is barely detectable by the coherent backscattering or the coherent
forward scattering experiment. This loss-insensitive point is located approximately at the turning
points of the results for $l_s$. In the dips and the peaks of $l_s$ loss and
gain is most efficiently detected. Speaking in terms of the scattering
strength of the disordered sample, it is interesting, that while absorption
reduces the resonators influence in general and thus leads to the reduction of
the scattering strength in the Mie resonance position, it leads to a
remarkable increase of the scattering strength of the ensemble in all off-resonant cases,
e.g. for $\lambda\,=\,270.0\,nm$ to $\lambda\,=\,300.0\,nm$, $\lambda\,=\,320.0\,nm$ to
$\lambda\,=\,367.0\,nm$, $\lambda\,=\,390.0\,nm$ to $\lambda\,=\,470.0\,nm$,
and for $\lambda\,=\,515.0\,nm$ to $\lambda\,=\,580.0\,nm$. This effect is
pronounced for monodisperse samples.

\begin{figure}[t]
\vspace*{0cm} \hspace*{-0.1cm}\rotatebox{0}{\scalebox{0.36}{\includegraphics[clip]{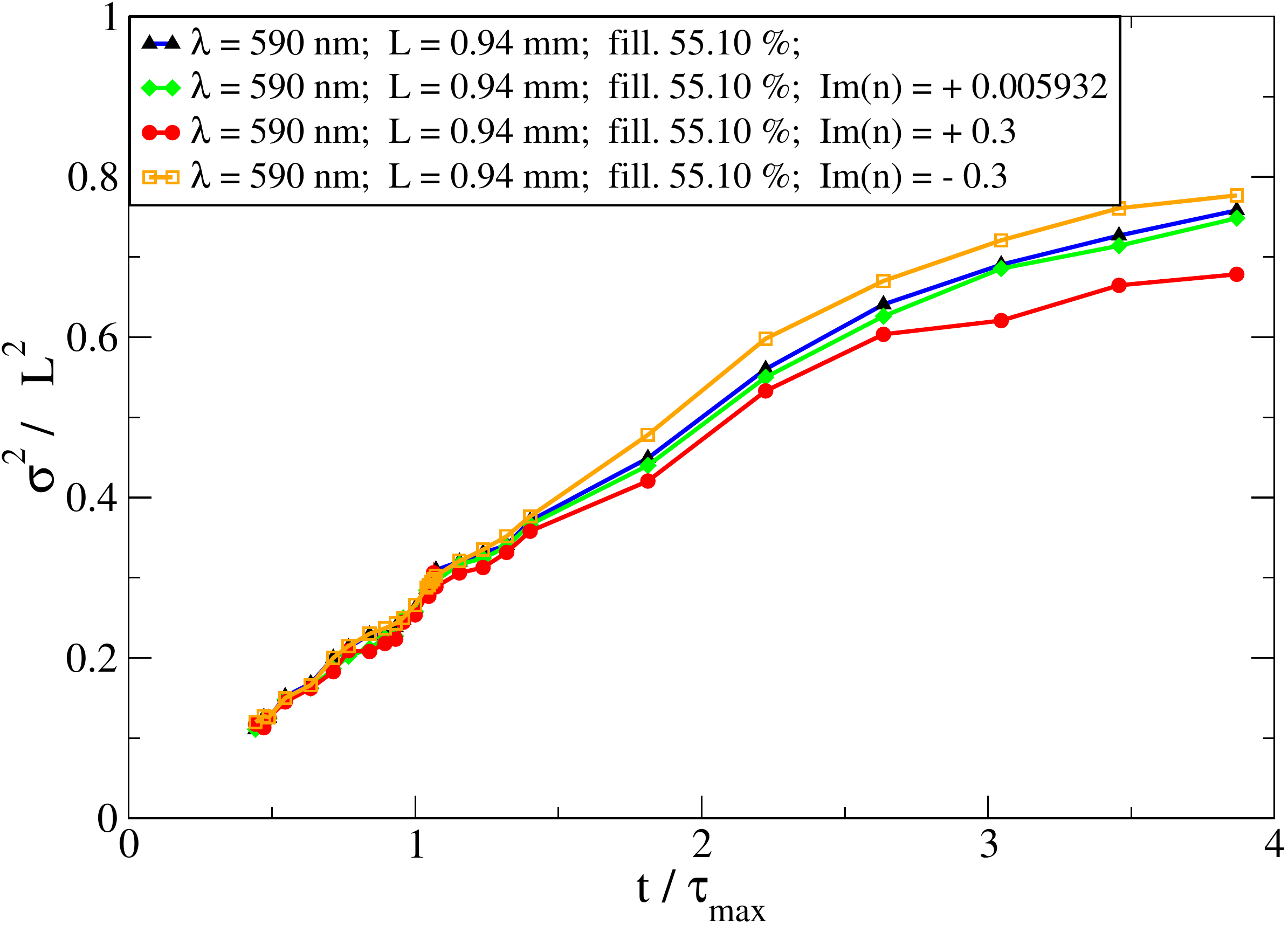}}}
\caption{Temporal evolution of $\sigma^2(t)/L^2$, the cross section of the
  transmitted photon density (1/e). The samples
  extent is $L\,=\,0.94\,mm$, $n\,=\,2.7$, the samples filling is $55.10 \%$. The
  Gaussian distribution of scatterers is centered at the diameter of
  $2\,r_{scat}\,=\,245.0\,nm$, full width half maximum is $17.0\,nm$. The
    scatterers' background is air, $\epsilon_b\,=\,1$. The analysing  pulse is
  characterized by it's center, which is $\lambda\,=\,590.0\,nm$, full width
  half maximum is $250.0\,fs$, beam waist on the glass surface at focus is $100.0\,\mu m$, the
  pulse transition time is $\tau_{max}\,=\,2.42\,ns$, see
  Fig.(\ref{Setup}). The result is normalized to the peak transition time. We
  display the passive complex system (black), the same arrangement including
  the literature value for absorption of TiO$_2$ bulk, ${\rm Im}\,(n)\,=\,0.005932$ (green),
an experimentally relevant value for absorption of disordered random TiO$_2$
media ${\rm Im}\,(n)\,=\,+0.3$ (red), and the same scatterers arrangement with
refractive index $n\,=\,2.7$ as it is pumped and gain assumes a value of
${\rm Im}\,(n)\,=\,-0.3$ (yellow). For all parameters a plateau in the temporal
evolution of $\sigma^2(t)/L^2$ is found as a characteristics of the complex
medium in the long time limit. Whereas for gain the spreading of the cross-section overall is
increased, absorption generally seems to lead to an earlier inset of the
plateau effect. Results for the passive system (black), ${\rm Im}\,(n)\,=\,0.005932$ (green) and
${\rm Im}\,(n)\,=\,+0.3$ (red) are to be compared to results of the scattering mean
free path $l_s$ Fig. (\ref{LABEL0}), and to the diffusion
coefficient $D$, Fig. (\ref{D_GAUSS}), at $\lambda\,=\,590.0\,nm$.} 
\label{Sigma2dL2}
\end{figure}

\begin{figure}[t]
\vspace*{-0.0cm} \hspace*{-0.2cm}\rotatebox{0}{\scalebox{0.35}{\includegraphics[clip]{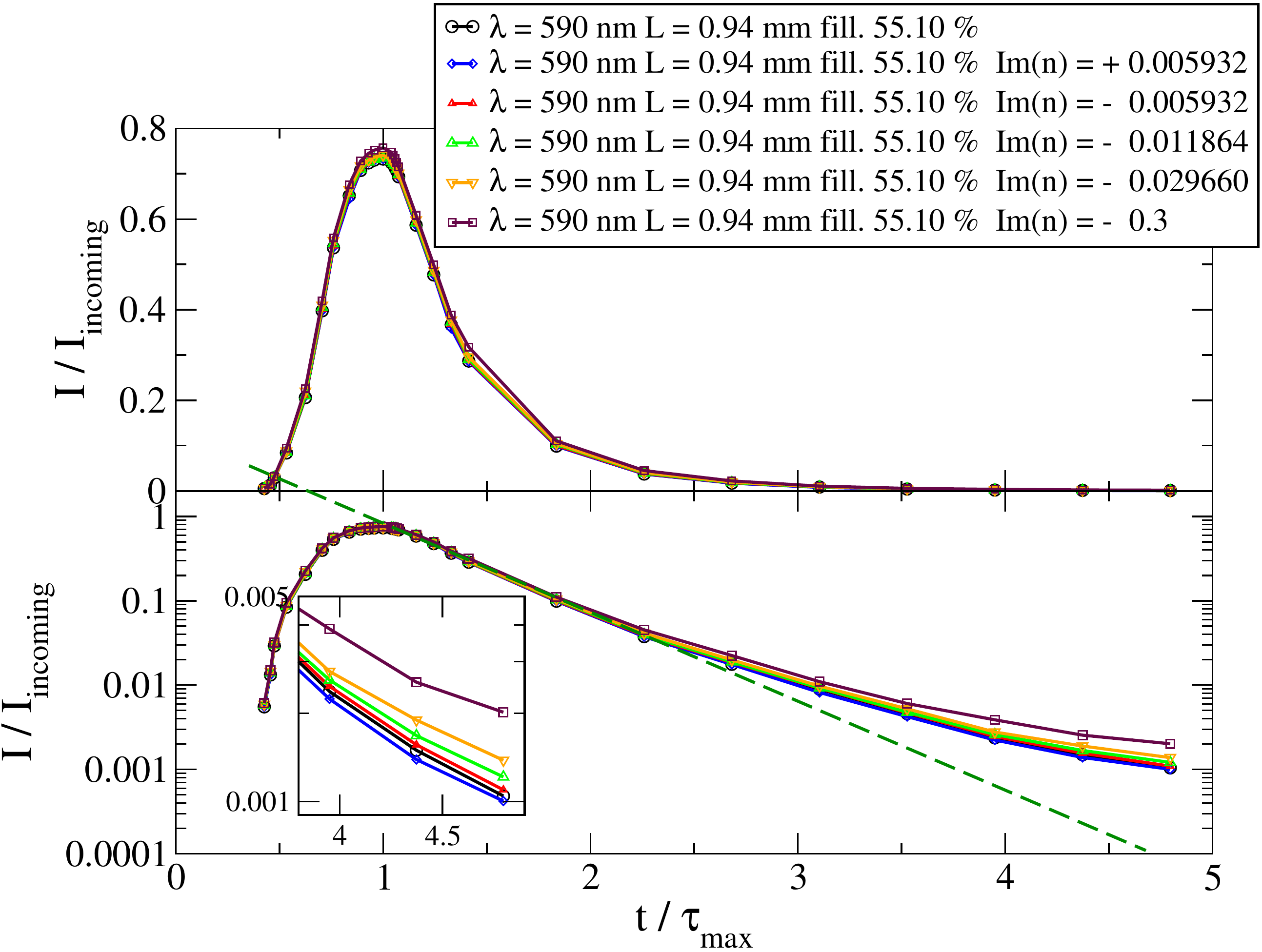}}}
\caption{Temporal evolution of intensity transmission $I/I_{incoming}$. $I$ is the time-dependent, surface- and
     angle-integrated count rate at the detector. $I_{incoming}$ is the time-, surface- and angle-integrated count rate at the detector. The
  green line marks the behavior of purely diffusive transport, the deviation
  to coherent transport, see black line and circles, in the log-plot is
  significant. The result is normalized to the peak transition time
  $\tau_{max}\,=\, 2.42\,ns$. The bottom panel shows the log-plot of the result
where the green dashed line marks the diffusive limit. We display results
for the passive case (black), and the same arrangement including
  the literature value for absorption of TiO$_2$ bulk, ${\rm Im}\,(n)\,=\,0.005932$,
(blue). One finds that absorption is pulling the result in the long time limit
towards the purely
diffusive case. Gain on the other hand, ${\rm Im}\,(n)\,=\,- 0.005932$,
${\rm Im}\,(n)\,=\,- 0.005932$ (red), ${\rm Im}\,(n)\,=\,- 0.011864$ (green),
${\rm Im}\,(n)\,=\,- 0.029660$ (orange), and ${\rm Im}\,(n)\,=\,- 0.3$ (maroon), yields a
lift of the result in the long time limit.}
\label{Label4}
\end{figure}

The diffusion coefficient $D$ as it is formally
derived in Eq. (\ref{D_omega_full}) can in principal be a complex quantity where
the imaginary part ${\rm Im} D$ becomes of physical relevance for pumped active
complex matter near or at the laser threshold. In coherent backscattering
experiments and in coherent forwardscattering experiments the real part of the
diffusion coefficient $Re D$ plays the crucial role Fig. (\ref{D_GAUSS}). $Re D$ is commonly
addressed in the literature as the diffusion constant $D$ and it can be implicitly measured.
For an absorbing polydisperse sample, $\tau_a\,=\,0.9\,ns$ and
${\rm Im}(n)\,=\,0.3$, we find that almost any Mie
characteristics is washed out due to the absorptive character of the single
scatterer. A destructive
interplay between absorption characteristics and the resonator properties is
developed all over the electromagnetic excitation spectrum. Extremely
pronounced is the difference in the magnitude of the diffusion coefficient $D$
for wavelengths larger than $\lambda\,=\,700.0\,nm$. When the filling is
enhanced from $48.83\,\%$ to $55.10\,\%$ the diffusion coefficient $D$  at
$\lambda\,=\,800.0\,nm$ is reduced from $85\,m^2/s$ to $25\,m^2/s$ so
approximately to $30\,\%$. Is shall be pointed out here that there is no transition
to an ordered sample considered and all results are derived for homogeneous
disordered but polydisperse samples of the mentioned Gaussian distribution of
scatterer radii. We display one of the crossover points in the inset of
Fig. (\ref{D_GAUSS}) the diffusion coefficient $D$ which is in the vicinity of the wavelength
$\lambda\,=\,583.0\,nm$.

When we discuss these results for $l_s$ and $D$, in terms of the Ioffe-Regel
criterium with the benchmark of strong localization of light, we find $kl\,<\,1$
could be experimentally derived for the monodisperse case,
see Fig. (\ref{LABEL0}), e.g. in the Mie resonance at $\lambda\,=\,490.0\,nm$,
whereas for the polydisperse ensemble $kl\,=\,1.67$ the condition is not
strictly fulfilled and interference effects will play a subtle role. It can thus be concluded that the probability to find Anderson localized photons will be
enhanced in the monodisperse ensemble where a factor of 2 enhancement with
respect to the incoming intensity could be detected. The increase of the volume filling
fraction for polydisperse samples will only have a moderate impact in the search for Anderson localized photons, whereas
absorption, which is enhanced in disordered granular media as compared to the
bulk case, is a crucial and limiting factor, see Fig. (\ref{L_S_GAUSS}). Thus it
is important to find a systematic theoretical method to distinguish micro- and
macroscopic structural effects in the signal which is a mix of both.

\subsection{Temporal evolution of the transmission cross section and the transmitted intensity}

In section {\it Quantum Field Theory for Multiple Scattering of Photons} \ref{QFT} we presented the theory to study the propagation and the
localization of a laser pulse through a disordered ensemble of complex
scatterers. The solution of this framework in the sense of the simulation of
detectable scattered intensity is not restricted to a dimension or a direction.
Transmission optical coherence tomography based
measurements of optical material properties
 is one experimental platform for our methodology \cite{Hee,Zaccanti,Trull}. Here we present results for the temporal evolution of the
transmission cross section $\sigma$, and the mean square width $\sigma^2$
respectively, Fig. (\ref{Sigma2dL2}), and the temporal
dependency of the transmitted photonic intensity,
Fig. (\ref{Label4}). Our
results are derived for homogeneous disordered samples with a Gaussian
distribution of scatterer diameters centered at $d\,=\,245.0\,nm$ and
$55.10\,\%$ volume filling. This is the strongly scattering and strongly
disordered regime. We display results
for the excitation laser frequency of $\lambda\,=\,590.0\,nm$, where we know
$kl_s\,=\,3.62$ from previous results, see Fig. (\ref{L_S_GAUSS}). In terms of
the Ioffe-Regel criterium this case should be far off any case of the Anderson
localization regime, thus the aim of our considerations is to determine the
number of coherently interfering photons as the deviation from the purely
diffusive case. The mean square width \cite{Buehrer,Maret2012,NPHOT2013}

\begin{equation}
\sigma^2 (t)\,=\,\frac{\int r^2 P_{E}(r,t)\, d^2
  r}{\int P_{E} (r,t)\, d^2 r}
\end{equation}

is defined as the square of the up to the
full-width-half-maximum (FWHM) limit integrated area $\sigma$ 
of the transmitted transverse intensity distribution at the ensemble surface, see
Fig.(\ref{Setup}). We display this characteristics in Fig. (\ref{Sigma2dL2})
normalized to the square of the samples length $L^2$.

We find that the transmission cross section, the mean square width,
Fig. (\ref{Sigma2dL2}), is definitely depending on absorption and gain
of the complex random medium, and it is a very sensitive measure. In
Fig. (\ref{Sigma2dL2}) we display results for the temporal evolution of $\sigma^2/L^2$ for the wavelength of the incident
pulse of $\lambda\,=\,590.0\,nm$. The case of $\lambda\,=\,590.0\,nm$ is extremely close to the crossing point
or turning point of the scattering mean free path $l_s$,
Fig. (\ref{L_S_GAUSS}), and the corresponding diffusion coefficient $D$,
Fig. (\ref{D_GAUSS}). While the characteristics  $l_s$ and $D$ are almost insensitive to absorption
at $\lambda\,=\,590.0\,nm$, a clear deviation
of $\sigma^2/L^2$ for the case of ${\rm Im}(n)\,=\,+0.3$ from the
passive case of about $12\,\%$ is found in the temporal limit of
$t/\tau_{max}\,=\,4$. In Fig. (\ref{Label4}) we show the results for the
transmitted intensity $I/I_{incident}$ in the long time limit. The purely
diffusive case is marked in the log-plot as the dashed green line. It can be
concluded for the strictly passive case (black line) that the increase of the
transmission in the long-time limit in comparison to the diffusive case with
the exponential decay is the number of localized photons, that have been
multiply and coherently scattering and interfering in the disordered medium in
the sense of the maximally crossed processes, represented diagrammatically by
the Cooperon, see  Fig. (\ref{Setup}). It is further derived that absorption, incorporated by the literature value for bulk TiO$_2$, ${\rm
  Im}(n)\,=\,+0.005932$ (blue line), reduces the number of localized photons,
and the long time behavior of $P_E$ retrogrades towards the diffusive
limit. This result as such has been expected. By comparison of the relative
magnitude of the results for $\sigma^2/L^2$ with the transmitted intensity
$P_E$, Fig. (\ref{Label4}), it is however interesting to note, that the plateau effect for the case of absorption is enhanced, precisely it shows an earlier onset. Thus the plateau as such seems not to be
a signature of localization, however the magnitude of the deviation of the
transmitted intensity $P_E$ from the diffusive limit can be interpreted as a
sign of enhanced coherent multiple scattering and thus as an enhancement of
interference effects in principle.
The influence of the single Mie scatterer is noteworthy when we discuss the
influence of gain. Gain as the negative imaginary part of the
complex refractive index and as a negative part of the complex permittivity is a
microscopic material characteristics equivalent to absorption. Whereas
absorption however is a microscopic interaction where the life-times of
light-matter bound states in first instance do not play a crucial role, this is different for the case of gain. Gain is
achieved by an enhanced life-time of light matter bound states leading to
an increase of the photon number in the incident wavelength. Gain and
absorption are as such in our theory properties of the single complex Mie
scatterer, and the microscopic materials characteristics is interacting with
the resonator. In the case of strong external laser pulses the local density
of states and thus the refractive index of TiO$_2$ bulk can be shifted and
this effect will lead to a shift of the overall characteristics of the
disordered granular medium $l_s$, Fig. (\ref{L_S_GAUSS}), and $D$, Fig. (\ref{D_GAUSS}). When non-linear effects, 
e.g. higher-order harmonics of the incident wavelength $\omega$, play a role,
this processes might contribute to an enhancement of multiple scattering and of
interference effects. These enhancements can result in time of flight
experiments is a variety of observations, such as off-centered peak in the
integrated spectrum. In this article we discuss gain in the central
wavelength of the incident pulse, which can originate from light-matter bound
states of higher harmonics, so non-linear effects play a role even though the
system is far below any laser threshold. The gain is visible on the one hand
side as a delayed onset of the plateau of $\sigma^2/L^2$ on the one hand,
see Fig. (\ref{Sigma2dL2}) (yellow line), on the other hand we find an
increase in the transmitted coherent photon intensity in the long-time limit,
Fig. (\ref{Label4}). The variation of $P_E$ due to enhanced multiple scattering
and interference effects due to an increased filling of due to enhanced
resonator properties of the geometrical scatterer can thus be well
determined and distinguished by comparing the characteristics of $l_S$, $D$,
$\sigma^2/L^2$ and $P_E$ from effects incorporating gain and absorption.

\section{Conclusions}

We have presented in this article an innovative method for characterizing disordered
complex random media based on a quantum-field theoretical approach, the
Vollhardt-W\"olfle theory, for photonic transport including interference
effects in multiple scattering processes. Our theory incorporates the
Ward-Takahashi identity for photonic transport and thus enables us to
determine self-consistent results for the material characteristics of
disordered granular complex media. We solve the theory in three dimensions space- and
time-dependent with a weighted essentially non-oscillatory solver method
(WENO). The solution with the WENO solver enables us to determine results
for ultrashort analyzing pulses and pump-probe experiments, since we can deal
with highly non-linear processes and discontinuities. We have presented here
a systematic study of the scattering mean free path $l_s$ and the diffusion
coefficient $D$. These characteristics can be compared directly to experimental results
derived in a coherent backscattering experiment. They show that the resonator
characteristics, e.g. the Mie resonance, plays a crucial role which is even
more important, when non-linear effects, gain and absorption, are minimized by
the choice of the scattering matter. It has also been shown that a
polydispersity of the scatterers reduces the probability to reach an Anderson
transition with transport of light. The incorporation of gain and absorption
reveals, that all materials characteristics are very sensitive to such
properties of complex matter. It has been shown that so called absorption-free
measurements are bound to spectrally narrow areas where the resonator
characteristics of the single scatterer leads to a minimized sensitivity with
respect to a change of the complex refractive index and the complex
permittivity. Our results for ab initio simulations of time of flight
experiments yield the characteristics of the normalized transmission or
reflection cross
section and the absolute as well as the normalized number of coherently
scattered photons. We presented results  random mono- and polydisperse
ensembles of TiO$_2$ Mie scatterers in a transmission optical coherence
tomography setup. It has been demonstrated that these characteristics offer
an increased sensitivity to any microscopic and the macroscopic structural
modification compared to the coherent backscattering experiment. The
underlying theory paves a way towards the detection of subtle interference
effects due to multiple scattering events in OCT setups that
may lead to an increase of the sensitivity of OCT of orders of magnitude, and
furthermore it may  improve the analysis of other methods of advanced
   spectroscopy like DWS, DLS and QFS. This
effect can be enhanced by the scatterers resonances. We conclude that our
combinatory analysis of underlying transport theory and its results, the scattering mean free path, the diffusion constant, and the
derived characteristics in the temporal evolution is suitable to distinguish
between perfectly coherent multiple scattering and interferences and on the
other hand between influences of the complex random medium in the
full spectrum of the analysis. We provide a consistent method, which is able
to characterize disordered media in the weakly and the strongly scattering
regime, the approach is suitable to incorporate ultrashort and intense light
pulses and the resulting subtle local and non-local light-matter interactions
on a broad temporal range. The method is ab initio not limited to light, it
can be performed with the full spectrum of electromagnetic excitations and 
it can be transferred to any other type of wave propagation as matter waves
and sound. It will be subject of subsequent work
to investigate further influences of multiple scattering and higher-order
non-linear effects in ensembles of clusters and composite scatterers such as
shells, as well as of macromolecules that can be
 random or quasi-ordered and may form so called {\it meta glasses}.\\

{\bf Authors contributions:} All authors were equally involved in the preparation of the manuscript.
All the authors have read and approved the final manuscript.\\
%
%

{\bf Acknowledgment:} The authors thank S. Fishman, L. Sanchez-Palencia, B. A. van Tiggelen, R. v. Baltz, W. B\"uhrer, G. Maret, K. Busch and J. Kroha for highly valuable discussions.

\bibliography{basename of .bib file}

\begin{thebibliography}{}
%

\bibitem{Akkermans} E. Akkermans, P. E. Wolf, R. Maynard, {
Phys. Rev. Lett.} {\bf 56} 14, 1471-1474 (1986).

\bibitem{Pecora} R. Pecola, { J. Chem. Phys.} {\bf 40}, 1604
  (1964).

\bibitem{Goodman} J. W. Goodman, { JOSA} {\bf 66}, 11, 1145-1150 (1976).

\bibitem{Provencher} S. W. Provencher, { Makromol. Chem.} {\bf 180},
  201-209 (1979). 

\bibitem{Schurtenberger} C. Urban, P. Schurtenberger, { J. Colloid. Interface
    Sci.}  {\bf 207} (1), 150-158 (1998). 


\bibitem{Scheffold} I. Block, F. Scheffold, { Rev. Sci. Instruments.} {\bf
    81}, 12, 123107-123107-7 (2010). 


\bibitem{Baravian} C. Baravian, F. Caton, J. Dillet, J. Mougel,
Phys. Rev. E {\bf 71}, 066603 (2005).


\bibitem{Pine} D. J. Pine, D. A. Weitz, P. M. Chaikin, E. Herbolzheimer, {
    Phys. Rev. Lett.} {\bf 60} 12 1134-1137 (1988).


\bibitem{OCT} D. Huang, E. A. Swanson, C. P. Lin, J. S. Schuman,
  W. G. Stinson, W. Chang, M.  R. Hee, T. Flotte, K. Gregory, C. A. Puliafito,
  J. G. Fujimoto, { Science} {\bf 254}
  (5035) 1178-1181 (1991). 


\bibitem{Drex} A. F. Fercher, W. Drexler, C. K. Hitzenberger,
  T. Lasser, Rep. Prog. Phys. {\bf 66}, 239-303 (2003).

\bibitem{OCTIEEE} J. M. Schmitt, { IEEE Journal of Selected Topics
    in Quantum Electronics}, {\bf 5}, No. 4, 1205-1214 (1999).

\bibitem{Zhang} A. Zhang, Q. Zhang, Ch.-L. Chen, R. K. Wang,
  Journal of Biomedical Optics {\bf 20}(10), 100901 (2015).

\bibitem{Zhou} K. C. Zhou, R. Qian, S. Degan, S. Farsiu, J. A. Izatt, Nature
  Phot. doi:10.1038/s41566-019-0508-1 (2019).

\bibitem{Dogariu2} A. Podoleanu, I. Charalambous, L. Plesea, A. Dogariu,
  R. Rosen, Phys. Med. Biol. {\bf 49}, 1277 (2004).



\bibitem{John1} S. John, {Phys. Rev. Lett.} {\bf 58}, 23, 2486-2489 (1987).


\bibitem{John2} F. C. MacKintosh, S. John, {Phys. Rev. B} {\bf 37}, 4,
  1884-1897 (1988).

\bibitem{deBoer} J. F. de Boer, T. E. Milner, J. S. Nelson,
  Opt. Lett. {\bf 24}, 300-302 (1999).

\bibitem{Yamanari} M. Yamanari, S. Tsuda, T. Kokubun, Y. Shiga, K. Omodaka, N.
Aizawa, Y. Yokoyama, N. Himori, S. Kunimatsu-Sanuki, K.
Maruyama, H. Kunikata, T. Nakazawa,  Biomed. Opt.
Express {\bf 7}, 3551-3573 (2016).

\bibitem{Gosh} N. Gosh, M. F. G. Wood, A. Vitkin,  J. Biomed. Opt. {\bf 13}, 044036 (2008).

\bibitem{Gompf1} B. Gompf, M. Gill, M. Dressel, A. Berrier, JOSA {\bf 35} (2)
  301-308 (2018). 

\bibitem{Serpo1}  H. Roychowdhury, S. A. Ponomarenko, E. Wolf,  {J. Mod. Opt.}
  52(11), 1611-1618 (2005).

\bibitem{Serpo2} S.A. Ponomarenko, Opt. Lett. {\bf 40}(4), 566-568 (2015).

\bibitem{Dogariu} O. Korotkova, M. Salem, A. Dogariu, E. Wolf, Waves in Random
  Complex Media 15(3), 353-364 (2005).

 \bibitem{Gompf} I. Voloshenko, B. Gompf, A. Berrier, M. Dressel,
  G. Schnoering, M. Rommel, J. Weis, Appl. Phys. Lett. {\bf 115}, 063106 doi:10.1063/1.5094409 (2019).


\bibitem{Vellekop_Aegerterimaging} I. M. Vellekop, C. M. Aegerter, { Optics
    Letters} {\bf 35} 8, 1245 (2010).

\bibitem{Mosk} A. P. Mosk, A. Lagendijk, G. Lerosey, M. Fink, { Nature
    Photonics} {\bf 6}, 283-292 (2012).

\bibitem{Lippok} N. Lippok, M. Siddiqui, B. J. Vakoc, B. E. Bouma,
  Phys. Rev. Applied {\bf  11}, 014018 (2019).



\bibitem{Isichenko} M.B. Isichenko, { Rev. Mod. Phys.}, {\bf 64}, 4, 961
  (1992).



\bibitem{Mandel}  Z. Y. Ou, L. J. Wang, X. Y. Zou, L. Mandel,
  Phys. Rev. A {\bf 41}, 1597 (1990).

\bibitem{Ou}  M. V. Chekhova, Z. Y. Ou, Adv. Opt. Photon. 8, 104
  (2016).

\bibitem{Teich1} A. F. Abouraddy, M. B. Nasr, B. E. A. Saleh,
  A. V. Sergienko, M. C. Teich, Phys. Rev. A {\bf 65}, 053817 (2002).

\bibitem{Teich2} M. B. Nasr, B. E. A. Saleh, A. V. Sergienko, M. C. Teich,
  Optics Express {\bf 12} (7), 1353-1362 (2004).

\bibitem{Teich3} M. B. Nasr, B. E. A. Saleh, A. V. Sergienko, M. C. Teich,
  Phys. Rev. Lett. {\bf 91}, 083601 (2003).

\bibitem{Amon} A. Amon, A. Mikhailovskaya, J. Crassous,
Review of Scientific Instruments {\bf 88}, 051804 (2017).

\bibitem{Wuensche}  S. Fuchs, C. R\"odel, A. Blinne, U. Zastrau, M. W\"unsche,
  V. Hilbert, L. Glaser, J. Viefhaus, E. Frumker, P. Corkum, E.
F\"orster, and G. G. Paulus, Sci. Rep. {\bf 6}, 20658 (2016).

\bibitem{Torres} A. Valles, G. Jimenez, L. J. Salazar-Serrano, J. P. Torres,
  Phys. Rev  A {\bf 97}, 023824 (2018).


\bibitem{Novitsky} D.V. Novitsky, EPL, {\bf 99} 44001 (2012).


\bibitem{Tearney} G. J. Tearney, M. E. Brezinski, J. F. Southern, B. E. Bouma, M. R. Hee, and J. G. Fujimoto, Opt. Lett. {\bf 20}, 2258-2260 (1995).

\bibitem{Knuttel} A. Knuttel, M. Boehlau-Godau, J. Biomed. Opt. {\bf 5}, 83-92 (2000).

\bibitem{Zvyagin} A. V. Zvyagin, K. K. M. B. D. Silva, S. A. Alexandrov,
  T. R. Hillman, J. J. Armstrong, T. Tsuzuki, D.D. Sampson, Opt. Express {\bf
    11}, 3503-3517 (2003).


\bibitem{Alfano} B. B. Das, F. Liu, R. R. Alfano, Rep. Prog. Phys.
{\bf 60}, 227-292 (1997). 

\bibitem{Nieuwenhuizen} M. C. W. van Rossum, Th. M. Nieuwenhuizen,
Rev. Mod. Phys. {\bf 71}, 313 (1999).

\bibitem{Rotter2} R. Savo, R. Pierrat, U. Najar, R. Carminati, S. Rotter,
  S. Gigan, Science, {\bf 358} (6364), 765-768 (2017).

\bibitem{Rotter} S. Rotter, S. Gigan, Rev. Mod. Phys. {\bf 89}, 015005 (2017).

\bibitem{Mycek} K. Vishwanath, B. Pogue, M.-A. Mycek, Phys. Med. Biol. {\bf
    47}, 3387-3405 (2002).

\bibitem{CaoPNAS} B. Redding, A. Cerjan, X. Huang, M. Larry Lee,
  A. D. Stone, M. A. Choma, H. Cao,  PNAS {\bf 112} (5) 1304-1309 (2015).

\bibitem{Chu} O. Liba, M. D. Lew, E. D. SoRelle, R. Dutta, D. Sen, D. M. Moshfeghi, S. Chu, A. de la Zerda, Nat. Comm. 8:15845 (2017).

\bibitem{Sun} T.-M. Sun, C.-S. Wang, C.-S. Liao, S.-Y. Lin, P. Perumal,
  C.-W. Chiang, Y.-F. Chen, ACS Nano {\bf 9}(12), 12436-12441 (2015).

\bibitem{Erden} K. V. Chellappan, E. Erden, H. Urey, Appl. Opt. {\bf 49}, F79-F98 (2010).


\bibitem{Boccara}  A. Badon, D. Li, G. Lerosey,
  A. C. Boccara, M. Fink, A. Aubry, Sci. Adv. 2016; 2 : e1600370 (2016).



\bibitem{Vignolini} G. Jacucci, J. Bertolotti, S. Vignolini,
Adv. Optical Mater. 1900980 (2019).

\bibitem{Garcia}  P.D. Garcia, R. Sapienza, A. Blanco, C. Lopez,
  Adv. Mater. 19:2597-2602 (2007).

\bibitem{Scheffold2} N. Senbil, M. Gruber, C. Zhang, M. Fuchs, F. Scheffold,
  Phys. Rev. Lett. {\bf 122}, 108002 (2019).

\bibitem{FrankPRB2006} R. Frank, A. Lubatsch, J. Kroha, { Phys. Rev. B} {\bf
    73}, 245107 (2006).

\bibitem{Frank2011} R. Frank, A. Lubatsch, { Phys. Rev. A} {\bf 84}, 013814
  (2011).

\bibitem{Shu} Chi-Wang Shu, 
{ NASA/CR-97-206253. ICASE
  Report} {\bf 97-65}, (1997).




\bibitem{Harten} A. Harten, B. Engquist, S. Osher, S. R. Chakravarthy,
{ Journal of Computational Physics} {\bf 71}, 231-303 (1987).

\bibitem{Osher}  C.-W. Shu, S. Osher, 
{ Journal of Computational Physics} {\bf 77}, 439-471 (1988).

\bibitem{Liu} X. Liu, S. Osher, T. Chan, 
  { Journal of Computational Physics} {\bf 115}, 200-212 (1994).

  \bibitem{Shu2020} Y.-T. Zhang, J. Shi, C.-W. Shu, Y. Zhou, Phys. Rev E {\bf 68}, 046709
(2003).

\bibitem{Lax} P. D. Lax, R. D. Richtmyer, 
{ Comm. Pure Appl. Math.} {\bf 9}, 267-293 (1956). 

\bibitem{Warburton} J. S. Hesthaven, T. Warburton, 
{ Philosophical Transactions of the Royal Society A: Mathematical,
  Physical and Engineering Sciences}, {\bf 362} (1816), 493-524 (2004).


\bibitem{VW1} D. Vollhardt, P. W\"olfle, { Phys. Rev. Lett.} {\bf 45}, 842
  (1980); 

\bibitem{VW2} D. Vollhardt, P. W\"olfle, {Phys. Rev. B} {\bf 22}, 4666 (1980).



\bibitem{Anderson} P. W. Anderson, { Phys. Rev.} {\bf 109}, 1492 (1958).



\bibitem{Chabanov} A. A. Chabanov, M. Stoytchev, A. Z. Genack, Nature {\bf 404}, 850–853 (2000).

\bibitem{PhysicsToday} A. Lagendijk, B. v. Tiggelen, D. S. Wiersma, {
    Physics Today}  {\bf 62} (8), 24 (2009). 

\bibitem{Fishman} T. Schwartz, G. Bartal, S. Fishman, M. Segev, { Nature}
  {\bf 446}, 52-55 (2007).

\bibitem{SegevChristodoulides} J. W. Fleischer, M. Segev, N. K. Efremidis,
  D. N. Christodoulides, { Nature} {\bf 422}, 147-150 (2003). 

\bibitem{Genack2}  N. Garcia, A. Z. Genack,  { Phys. Rev. Lett.} {\bf 66}
  1850-1853 (1991).

\bibitem{Shapiro1}  P. Henseler, J. Kroha, B. Shapiro
{ Phys. Rev. B} {\bf 78}, 235116 (2008).

\bibitem{Shapiro2} C. A. M\"uller, D. Delande, B. Shapiro, {Phys. Rev. A} {\bf
    94}, 033615 (2016).

\bibitem{SAspect} J. Billy, V. Josse, Z. Zuo, A. Bernard, B. Hambrecht,
  P. Lugan, D. Clement, L. Sanchez-Palencia, O. Bouyer, A. Aspect, {
  Nature} {\bf 453} (7197), 891-894 (2008).


\bibitem{HuS} H. Hu, A. Strybulevych, J.H. Page, S.E. Skipetrov,
B.A. van Tiggelen, Nat. Phys. 4, 945 (2008).

\bibitem{Maret2012} T. Sperling, W. B\"uhrer, C. M. Aegerter, G. Maret, {
    Nature Photon.} {\bf 7}, 48-52 (2013).

\bibitem{WS} F. Scheffold, D. Wiersma, { Nature Photon.} {\bf 7} (12), 934
  doi: 10.1038/nphoton.2013.210 (2013).

\bibitem{NPHOT2013} G. Maret, T. Sperling, W. B\"uhrer, A. Lubatsch, R. Frank,
  C.M. Aegerter, { Nature Photon.} {\bf 7} (12), 934-935
  doi:10.1038/nphoton.2013.281 (2013).

\bibitem{Sanchez2019} J. Richard, L.-K. Lim, V. Denechaud, V. V. Volchkov,
  B. Lecoutre, M. Mukhtar, F. Jendrzejewski, A. Aspect, A. Signoles,
  L. Sanchez-Palencia, V. Josse, Phys. Rev. Lett. {\bf 122}, 100403 (2019).




\bibitem{Vardeny} Z. V. Vardeny, A. Nahata, A. Agrawal, { Nature Phot.}
  {\bf 7}, 177-187 (2013).


\bibitem{Genack} J.I. Gersten, D.A. Weitz, T.J. Gramila, A. Z. Genack, {
    Phys. Rev. B} {\bf 22}, 10, 4562-4571 (1980). 


\bibitem{DUALSYM} J. C. J. Paasschens, T. Sh. Misirpashaev,
  C. W. J. Beenakker, { Phys. Rev. B} {\bf 54}, 11887 (1996).

\bibitem{Evans} C. C. Evans, J. D. B. Bradley,
E. A. Marti-Panameno, E. Mazur, { Optics Express} {\bf 20}, 3,  3118 (2012).

\bibitem{Chakravarthy} G. Chakravarthy, S. R. Allam, A. Sharan, {
    J. Nonlinear Optic. Phys. Mat.} {\bf 25}, 2, 1650019 (2016).


\bibitem{Sanchez-PLewenstein} L. Sanchez-Palencia, M. Lewenstein,
{ Nat. Phys.} {\bf 6}, 81 (2010).

\bibitem{Mie} G. Mie, { Ann. Phys.} (Leipzig) {\bf 330}, 377 (1908).

\bibitem{Lubatsch05} A. Lubatsch, J. Kroha, K. Busch, { Phys. Rev. B} {\bf
    71}, 184201 (2005).



\bibitem{Wegener} M. Wegener, { Extreme Nonliner Optics}, ISBN 3-540-22291-X,
  Springer (2004).

\bibitem{Bohren}  G.F. Bohren, D.R. Huffman, { Absorption and scattering
  of light by small particles}. John Wiley$\&$Sons (1983).

\bibitem{Stoerzer1} M. St\"orzer, C. M. Aegerter, G. Maret, {
    Phys. Rev. Lett.} {\bf 96}, 063904 (2006).

\bibitem{Stoerzer2} M. St\"orzer, C. M. Aegerter, G. Maret, {
    Phys. Rev. E} {\bf 73}, 065602(R) (2006).

\bibitem{Buehrer} W. B\"uhrer, { Anderson Localization of Light in the
    Presence of Non-linear Effects}, Dissertation,
  http://nbn-resolving.de/urn:nbn:de:bsz:352-207872 (2012).

\bibitem{WiersmaPRL2007} R. Sapienza, P.D. Garcia, J. Bertolotti, M. D. Martin,
  A. Blanco, L. Vina, C. Lopez, D.S. Wiersma, { Phys. Rev. Lett.} {\bf 99}
  233902 (2007).

\bibitem{WiersmaPRA2008} P. D. Garcia, R. Sapienza, J. Bertolotti,
  M. D. Martin, A. Blanco, A. Altube, L. Vina, D. S. Wiersma, C. Lopez,
  { Phys. Rev. A}  {\bf 78}, 023823 (2008).

\bibitem{WiersmaLevi2012} M. Burresi, V.  Radhalakshmi, R. Savo,
  J. Bertolotti, K. Vynck, D. S. Wiersma, { Phys. Rev. Lett.} {\bf 108},
  110604 (2012).

\bibitem{WiersmaFibonacci2005} M. Ghulinyan, C. J. Olton, L. Dal Negro,
  L. Pavesi, R. Sapienza, M. Colocci, D.S. Wiersma, { Phys. Rev. B} {\bf
    71} 09204 (2005).

\bibitem{WiersmaLAGPRE1996} D.S. Wiersma, A. Lagendijk, { Phys. Rev. E} {\bf
    54}, 4256-4265 (1996).

\bibitem{Kubo1} R. Kubo, 
J. Phys. Soc. Japan {\bf 12}, 570 (1957). See also R.
Kubo, in Lectures in Theoretical Physics, W. E. Brittin and L.
G. Dunham, Eds. (Interscience Publishers, Inc. ,
York, 1959), Vol. I, p. 120.



\bibitem{Peterson} R. L. Peterson, 
{Rev. Mod. Phys.} {\bf 39}, 1, 69-77 (1967).

\bibitem{LubatschEPJB} A. Lubatsch, R. Frank, Eur. Phys. J. B,
  {\bf 92}: 215 (2019).


\bibitem{LubatschSymmetry} A. Lubatsch, R. Frank, Symmetry {\bf 11}, 1246
  (2019).

  \bibitem{LubatschAPPLSCI2020} A. Lubatsch, R. Frank, Appl. Sci. {\bf 10}(5), 1836 (2020).


\bibitem{Reichl} L. E. Reichl, M. D. Porter, Phys. Rev. E, {\bf 97} (4)
  042206 (2018).


\bibitem{WARD}
J. C. Ward, 
{ Phys. Rev.} {\bf 78}, 2, 182 (1950).

\bibitem{TAKAHASHI}
 Y. Takahashi, 
{ Il Nuovo Cimento} Vol. VI, 2, 371-375 (2231 - 2235) (1957).

\bibitem{Kroha} J. Kroha, { Physica A: Statistical Mechanics and its
Applications},  Vol. 167, issue 1, 231-252 (1990).

\bibitem{OnsagerI} L. Onsager, 
Phys. Rev. 37, 405-426 (1931). 
\bibitem{OnsagerII} L. Onsager, Phys. Rev. 38, 2265 (1931).

\bibitem{Lagendijk_mie} K. L. van der Molen, P. Zijlstra, A. Lagendijk, A. P. Mosk, { Opt. Lett.} {\bf 31}, 1432 (2006).

\bibitem{NJP14} A. Lubatsch, R. Frank,  { New J. Phys.} {\bf 16},
083043 (2014).

\bibitem{SREP15} A. Lubatsch, R. Frank, { Scientific Reports}
  {\bf 5}, 17000 (2015).

\bibitem{ApplSci} A. Lubatsch, R. Frank {Appl. Sci.} {\bf 9}, 2477 (2019).

\bibitem{Busch} K. Busch, C.M. Soukoulis, { Phys. Rev. Lett.} {\bf
    75}, 3442 (1995); { Phys. Rev. B} {\bf 54}, 893 (1996).
 
\bibitem{Tip1} B. A. van Tiggelen, A. Lagendijk, M. P. van Albada, A. Tip,
{ Phys. Rev. B} {\bf 45}, 12233 (1992).

\bibitem{Tip2}  B. A. van Tiggelen, A. Lagendijk, A. Tip, {
    Phys. Rev. Lett.} {\bf 71}, 1284 (1993).

\bibitem{Hee} M. R. Hee, J. A. Izatt,
J. M. Jacobson, J. G. Fujimoto, Opt. Lett. {\bf 18}, (12), 950-952 (1993).

\bibitem{Zaccanti} G. Zaccanti, S. D. Bianco, and F.
  Martelli, Applied Optics {\bf 42}, (19) 4023 (2003).

\bibitem{Trull} A. K. Trull, J. van der Horst, J. G. Bijster, J. Kalkman,
  Optix Express {\bf
    23}, (26) 33550 (2015).



\end{thebibliography}

\end{document}